\documentclass[prd,aps,showpacs,tightenlines,nofootinbib,eqsecnum]{revtex4}
\usepackage{graphicx,color,amsmath,amsxtra}
\usepackage{amssymb}
\usepackage{enumerate}
\usepackage{hhline}
\usepackage{array}
\usepackage{tabularx}

\newcommand{\bear}{\begin{array}}  \newcommand{\eear}{\end{array}}
\newcommand{\bea}{\begin{eqnarray}}  \newcommand{\eea}{\end{eqnarray}}
\newcommand{\beq}{\begin{equation}}  \newcommand{\eeq}{\end{equation}}
\newcommand{\bef}{\begin{figure}}  \newcommand{\eef}{\end{figure}}
\newcommand{\bec}{\begin{center}}  \newcommand{\eec}{\end{center}}

\newcommand{\Eqn}[1]{&\hspace{-0.2em}#1\hspace{-0.2em}&}

\def\Vec#1{\mbox{\boldmath $#1$}}

\def\Lap{{\mathop{\Delta}\limits^{(3)}}}

\def\be{\begin{equation}}
\def\ee{\end{equation}}
\def\bea{\begin{eqnarray}}
\def\eea{\end{eqnarray}}
\def\beq{\begin{eqnarray}}
\def\eeq{\end{eqnarray}}

\def\nn{\nonumber \\}
\def\e{{\rm e}}

\begin{document}

\title{
Future of the universe in modified gravitational theories:
Approaching to the finite-time future singularity
}

\author{Kazuharu Bamba$^{1,2}$,
Shin'ichi Nojiri$^3$
and Sergei D. Odintsov$^4$\footnote{
Also at Lab. Fundam. Study, Tomsk State
Pedagogical University, Tomsk}}
\affiliation{
$^1$Department of Physics, National Tsing Hua University, Hsinchu, Taiwan 300\\
$^2$Kavli Institute for Theoretical Physics China, CAS, Beijing 100190, China\\
$^3$Department of Physics, Nagoya University, Nagoya 464-8602, Japan\\
$^4$Instituci\`{o} Catalana de Recerca i Estudis Avan\c{c}ats (ICREA)
and Institut de Ciencies de l'Espai (IEEC-CSIC),
Campus UAB, Facultat de Ciencies, Torre C5-Par-2a pl, E-08193 Bellaterra
(Barcelona), Spain
}


\begin{abstract}
We investigate the future evolution of the dark energy universe in modified
gravities including $F(R)$ gravity, string-inspired scalar-Gauss-Bonnet
and modified Gauss-Bonnet ones, and ideal fluid with the inhomogeneous
equation of state (EoS). Modified Friedmann-Robertson-Walker (FRW) dynamics
for all these theories may be presented in universal form by using the
effective ideal fluid with an inhomogeneous EoS without specifying its
explicit form.
We construct several examples of the modified gravity which produces
accelerating cosmologies ending at the finite-time future singularity of all
four known types by applying the reconstruction program.
Some scenarios to resolve the finite-time future singularity are presented.
Among these scenarios, the most natural one is related with additional
modification of the gravitational action in the early universe.
In addition, late-time cosmology in the non-minimal Maxwell-Einstein theory
is considered. We investigate the forms of the non-minimal gravitational
coupling which generates the finite-time future singularities and
the general conditions for this coupling in order that the finite-time
future singularities cannot emerge.
Furthermore, it is shown that the non-minimal gravitational coupling can
remove the finite-time future singularities or
make the singularity stronger (or weaker) in modified gravity.
\end{abstract}

\pacs{
11.25.-w, 95.36.+x, 98.80.Cq, 98.62.En
}

\maketitle

\section{Introduction \label{Sec1}}

Recent observational data strongly indicate the existence of the dark energy,
which generates the accelerating expansion of the present universe.
In particular, the five-year Wilkinson Microwave Anisotropy Probe (WMAP)
data~\cite{Komatsu:2008hk} give the bounds to the value of the equation of
state (EoS) parameter $w_{\rm DE}$, which is the ratio of
the pressure of the dark energy to the energy density of it,
in the range of $-1.11 < w_{\rm DE} < -0.86$.
This could be consistent if the dark energy is a cosmological
constant with $w_{\rm DE} = -1$ and therefore our universe seems to approach
to asymptotically de Sitter universe.
It is also believed that there existed the period of another accelerating
expansion of the universe, called inflation, in the early universe.
In many models of inflation, the accelerating expansion could be generated
by almost flat potential of scalar field(s), called inflaton.
Hence, in the period of the inflation the universe could be described by
(almost) de Sitter space.
Thus, there is the striking similarity between the very early and very late
universe.

Although both of the accelerating expansions seem to be de Sitter type,
the possibility that the current acceleration could be
quintessence type, in which $w_{\rm DE}>-1$, or phantom type, in which
$w_{\rm DE}<-1$, is not completely excluded.
Furthermore, even if the current accelerating universe is described by
$\Lambda$CDM epoch, it is quite possible that it may enter to
quintessence/phantom phase in future.
Similarly, even if the early-time inflation may be de Sitter type,
there may exist pre-inflationary stage in which the evolution of the
universe is different from de Sitter type, e.g., it could be
quintessence/phantom epoch.
It is often assumed that the early universe started from the singular
point often called Big Bang.
However,
if the current (or future) universe enters the quintessence/phantom stage,
it may evolve to the finite-time future singularity depending on the specific
model under consideration and the value of the effective EoS parameter.
This suggests that the same theory should describe the whole history of the
expansion of the universe.
To achieve this, two scalar fields, one field (inflaton) to
describe the inflation and another field to do the dark energy have often
been introduced.

In this paper, we consider another approach, which might be more natural than
introducing two scalar fields in order to unify the early-time inflation with
late-time acceleration: that is modified gravity (for a review,
see~\cite{review}). In this approach, one starts from some unknown
fundamental gravity. At the very early universe in which the curvature is very
large but quantum gravity effects may be neglected, the restricted specific
sector of such a theory predicts inflation.
In course of the evolution, the curvature decreases and the next-to-leading
terms become relevant at the intermediate universe
(the radiation/matter-dominated stage).
Note that the observational data seems to be consistent if the intermediate
universe could have been governed by the standard general relativity.
As the curvature becomes smaller, the universe enters the dark energy epoch
controlled by the different sector of the unknown fundamental gravitational
theory different from general relativity. The corresponding gravitational
terms are leading ones in comparison with the ones of general relativity
at the current curvature.
Hence, the evolution of the universe defines the modified gravitational theory
predicting its evolution at each stage~\cite{Lue:2003ky}. From other side,
the effective evolution of modified gravity is responsible for the history of
the expansion of the universe.
This also indicates that a right approach to the understanding of the
fundamental gravitational theory is to study the history of
the expansion of the universe, which will give the information about the
leading sectors of modified gravity at each epoch.
Moreover, the consistent examples of modified gravitational theories
passing the local tests and unifying the early-time inflation with
late-time acceleration have already been
constructed~\cite{Nojiri:2007cq, Nojiri:2007as}.
To understand such theories more clearly, it is reasonable to
investigate these theories at extreme situations, for instance, near to
singular points where some fundamental features of the theories may be
discovered.

Besides dark energy, there is another unknown component in the universe,
namely, dark matter. While dark energy has the large negative pressure,
the pressure of dark matter is negligible. In many scenarios, dark matter
could originate from particle physics, e.g., dark matter could be
a lightest supersymmetric (SUSY) particle, or particle coming from extra
dimensional theories.
There exists another kind of scenario in which dark matter would be
explained by the modification of the gravity, as in case of dark energy
considered in this paper.
There are many such scenarios like a modified Newtonian dynamics
(MOND)~\cite{Milgrom:1983ca}, Tensor-Vector-Scalar gravity
(TeVeS)~\cite{Bekenstein:2004ne}, and the generalized Einstein-Aether
theory~\cite{Zlosnik:2006zu} derived from the Einstein-Aether
theory~\cite{Eling:2004dk} and the equivalent one in non-relativistic
limit to MOND.
Dark matter might be explained by $F(R)$ gravity~\cite{Capozziello:2006ph},
where $F(R)$ is an arbitrary function of the scalar curvature $R$.
There is a scenario in which dark matter particles could be given
by the scalar mode in $F(R)$ gravity (the first paper in~\cite{review1}).

In the present paper, we study the future evolution of the dark energy epoch
for various modified gravities: $F(R)$ gravity, scalar-Gauss-Bonnet and
modified Gauss-Bonnet ones, and effective ideal fluid with the inhomogeneous
EoS, which includes the explicit dependence from Hubble rate and
curvature.
In particular, we are interested in the behavior of the accelerating
cosmological solutions in these theories when those solutions approach to the
finite-time future singularity. Of course, not all of these theories
predict such singularities. It depends on the value of the effective EoS
parameter and the structure of theories.
For instance, the expansion in the phantom phase which is not transient
predicts a future Big Rip singularity.

The paper is organized as follows.
In Sec.~II, we present the modified Friedmann-Robertson-Walker (FRW)
dynamics in a universal way. The following models are considered:
$F(R)$ gravity, scalar-Gauss-Bonnet and modified Gauss-Bonnet theories as
well as ideal fluid with the inhomogeneous EoS. It is indicated briefly
how the history of the expansion of the universe can be reconstructed for
such a universal formulation.
Sec.~\ref{Sec3} is devoted to the study of the finite-time future
singularities in $F(R)$ gravity. Using the reconstruction technique, we
present several examples which predict the accelerating FRW solutions ending
at the finite-time future singularity. It is demonstrated that not only the
Big Rip but other three types of the finite-time future singularities may
appear.
In Sec.~\ref{Sec4}, we discuss various scenarios to resolve the finite-time
future singularities. The most natural scenario is based on the additional
modification of the inhomogeneous EoS or the gravitational action by
the term which is not relevant currently. However, such a term which may be
relevant at very early or very late universe may resolve the finite-time
future singularities.
It is interesting that the presence of such a next-to-leading order term
does not conflict with the known local tests. Another scenario is related to
the account of quantum effects which become relevant near to the
finite-time future singularities.
Sec.~\ref{Sec5} is devoted to the construction of scalar-Gauss-Bonnet
and modified Gauss-Bonnet gravities which predict the late-time acceleration
ending in the finite-time future singularities.
Using the reconstruction technique, we present the corresponding effective
potentials.

In Secs.~\ref{Sec6} to \ref{Sec8}, we study some cosmological effects
in the non-minimal Maxwell-Einstein gravity with general gravitational
coupling. In Sec.~\ref{Sec6}, we describe our model and derive
the effective energy density and pressure of the universe.
In Sec.~\ref{Sec7}, we consider the finite-time future singularities
in non-minimal Maxwell-Einstein gravity.
We investigate the forms of the non-minimal gravitational coupling of the
electromagnetic field generating the finite-time future singularities
and the general conditions for the non-minimal gravitational coupling of
the electromagnetic field in order that the finite-time future singularities
cannot emerge.
Furthermore, in Sec.~\ref{Sec8} we consider the influence of non-minimal
gravitational coupling of the electromagnetic field on the finite-time
future singularities in modified gravity.
It is shown that a non-minimal gravitational coupling of the electromagnetic
field can remove the finite-time future singularities or make the singularity
stronger (or weaker).
Some summary and outlook are given in Sec.~\ref{Sec9}.
We use units in which $k_\mathrm{B} = c = \hbar = 1$ and denote the
gravitational constant $8 \pi G$ by ${\kappa}^2$, so that
${\kappa}^2 \equiv 8\pi/{M_{\mathrm{Pl}}}^2$, where
$M_{\mathrm{Pl}} = G^{-1/2} = 1.2 \times 10^{19}$GeV is the Planck mass.
Moreover, in terms of electromagnetism we adopt Heaviside-Lorentz units.

\section{Modified FRW dynamics and the reconstruction of the history of
the expansion of the universe \label{Sec2}}

In this section, we present the general point of view to the modification
of the FRW equations which may be caused by alternative gravity or ideal
fluid with the complicated EoS.
We consider $F(R)$ gravity, scalar-Gauss-Bonnet and modified Gauss-Bonnet
theories, and ideal fluid with the inhomogeneous EoS.
In addition, we briefly indicate how the history of the expansion of the
universe can be reconstructed through such a universal formulation.

The flat FRW space-time is described by the metric
\begin{eqnarray}
{ds}^2 =-{dt}^2 + a^2(t)d{\Vec{x}}^2\,,
\label{eq:KB1}
\end{eqnarray}
where $a(t)$ is the scale factor.
In the Einstein gravity, the FRW equations are given by
\be
\label{GBR18}
\rho = \frac{3}{\kappa^2}H^2\ ,\quad p = -\frac{1}{\kappa^2}\left(2\dot H + 3H^2\right)\ ,
\ee
where $H=\dot a/a$ is the Hubble parameter, a dot denotes a time
derivative, $\dot{~}=\partial/\partial t$, and $\rho$ and $p$ are
the energy density and pressure of the universe, respectively.
We have assumed the flat three-dimensional metric in accord with
observational data.
Let us consider any modified gravity (for a review, see~\cite{review})
like $F(R)$ gravity (for reviews, see~\cite{review, review1}), the
scalar-Gauss-Bonnet one, or
the modified Gauss-Bonnet one ($F({\cal G})$ gravity,
where ${\cal G}$ is the Gauss-Bonnet invariant given by
Eq.~(\ref{GB}) below).
In this case, the part of modified gravity may be formally included
into the total effective energy density and the pressure as in
Ref.~\cite{inh}, in which general inhomogeneous EoS fluid is introduced.
In this case, the modified FRW equations have the well-known form:
\be
\label{GBR18B}
\rho_{\rm eff} = \frac{3}{\kappa^2}H^2\ ,\quad
p_{\rm eff} = -\frac{1}{\kappa^2}\left(2\dot H + 3H^2\right)\ ,
\ee
where $\rho_{\rm eff}$ and $p_{\rm eff}$ are the effective energy density and
pressure of the universe, respectively.
Note that the contribution of (unusual) ideal fluid should also be included
in the left-hand side (l.h.s.) of the above modified FRW
equations~\cite{inh}.

$\rho_{\rm eff}$ and $p_{\rm eff}$ satisfy more general EoS as
\be
\label{EoS1}
p_{\rm eff} = - \rho_{\rm eff} + f\left(\rho_{\rm eff}\right)
+ G\left(H, \dot H, \ddot H, \cdots\right)\ ,
\ee
or even more complicated one.
The other point of view is possible: one can only keep the contribution of
matter in the energy density, while the gravitational modification should be
parameterized by the above function $G$.
We should note that $\rho_{\rm eff}$ and $p_{\rm eff}$ defined
in~(\ref{GBR18B}) satisfy the conservation law identically:
\be
\label{EoS0}
\dot \rho_{\rm eff} + 3H \left( \rho_{\rm eff} + p_{\rm eff} \right)=0 \ .
\ee

As an example, we may consider the $F(R)$ gravity~\cite{review,review1}
whose action is given by
\be
\label{fr1}
S_{F(R)}=\int d^4 x \sqrt{-g} \left\{\frac{F(R)}{2\kappa^2} + {\cal L}_m\right\}\ ,
\ee
where $F(R)$ is a proper function of the scalar curvature $R$ and
${\cal L}_m$ is the matter Lagrangian.
One may separate the modified part in $F(R)$ from the Einstein-Hilbert one as
\be
\label{fr2}
F(R)=R + f(R)\ .
\ee
In the FRW background with flat spatial part, $\rho_{\rm eff}$ and
$p_{\rm eff}$ are given by
\bea
\label{Cr4}
\rho_{\rm eff} &=& \frac{1}{\kappa^2}\left(-\frac{1}{2}f(R) + 3\left(H^2  + \dot H\right) f'(R)
 - 18 \left(4H^2 \dot H + H \ddot H\right)f''(R)\right) + \rho_{\rm matter}\ ,\\
\label{Cr4b}
p_{\rm eff} &=& \frac{1}{\kappa^2}\left(\frac{1}{2}f(R) - \left(3H^2 + \dot H \right)f'(R)
+ 6 \left(8H^2 \dot H + 4{\dot H}^2
+ 6 H \ddot H + \dddot H \right)f''(R) + 36\left(4H\dot H + \ddot H\right)^2f'''(R) \right) \nn
&& + p_{\rm matter}\ .
\eea
Here, $\rho_{\rm matter}$ and $p_{\rm matter}$ are the energy density and
pressure of the matter, respectively,
and the scalar curvature $R$ is given by $R=12H^2 + 6\dot H$.
If the matter has a constant EoS parameter $w$, Eq.~(\ref{EoS1}) has the
following form:
\bea
\label{EoS2}
\hspace{-5mm}
p_{\rm eff} &=& w \rho_{\rm eff} + G\left(H, \dot H, \ddot H, \cdots\right)\ ,\nn
\hspace{-5mm}
G\left(H, \dot H, \cdots\right) &=&
\frac{1}{\kappa^2}\left(\frac{1 + w}{2}f(R) - \left\{3\left(1+w\right)H^2
+ \left(1+3w\right) \dot H \right\}f'(R) \right. \nn
&& \left. + 6 \left\{ \left(8 + 12w\right) H^2 \dot H + 4{\dot H}^2
+ \left(6 +3 w\right) H \ddot H + \dddot H \right\}f''(R)
+ 36\left(4H\dot H + \ddot H\right)^2f'''(R) \right) \ .
\eea
Let us consider several examples.
In case of the model~\cite{NOprd} in which $f(R)$ is given by
\be
\label{EoSA1}
f(R)= - \frac{\alpha}{R} + \beta R^n\ ,
\ee
one finds
\bea
\label{EoSA2}
G\left(H, \dot H, \cdots\right) &=&
\frac{1}{\kappa^2}\left(\frac{1 + w}{2}\left( - \frac{\alpha}{R} + \beta R^n\right)
 - \left\{3\left(1+w\right)H^2
+ \left(1+3w\right) \dot H \right\}\left(\frac{\alpha}{R^2} + n\beta R^{n-1}\right) \right. \nn
&& + 6 \left\{ \left(8 + 12w\right) H^2 \dot H + 4{\dot H}^2
+ \left(6 +3 w\right) H \ddot H + \dddot H \right\}\left(- \frac{2}{R^3}
+ n(n-1)\beta R^{n-2} \right) \nn
&& \left. + 36\left(4H\dot H + \ddot H\right)^2 \left(\frac{6}{R^4} + n(n-1)(n-2) R^{n-3} \right) \right) \ .
\eea
For the Hu-Sawicki model~\cite{HS},
\be
\label{HS1}
f_{HS}(R)=-\frac{m^2 c_1 \left(R/m^2\right)^n}{c_2 \left(R/m^2\right)^n + 1}
= - \frac{m^2 c_1}{c_2}
+ \frac{m^2 c_1/c_2 }{c_2 \left(R/m^2\right)^n + 1}\ ,
\ee
we get
\bea
\label{EoSA3}
\hspace{-10mm}
&& G\left(H, \dot H, \cdots\right) =
\frac{1}{\kappa^2}\left[ - \frac{1 + w}{2}
\left(\frac{m^2 c_1 \left(R/m^2\right)^n}{c_2 \left(R/m^2\right)^n + 1}\right)
+ \left\{3\left(1+w\right)H^2
+ \left(1+3w\right) \dot H \right\}
\left( \frac{n c_1 \left(R/m^2\right)^{n-1} }{\left(c_2 \left(R/m^2\right)^n + 1\right)^2}\right)
\right. \nn
\hspace{-10mm}
&& \quad + 6 \left\{ \left(8 + 12w\right) H^2 \dot H + 4{\dot H}^2
+ \left( 6 +3 w\right) H \ddot H + \dddot H \right\}
\left(\frac{\frac{n(n-1) c_1}{m^2} \left(R/m^2\right)^{n-2} }{\left(c_2 \left(R/m^2\right)^n + 1\right)^2}
 - \frac{\frac{2 n^2 c_1 c_2}{m^2} \left(R/m^2\right)^{2n-2} }{\left(c_2 \left(R/m^2\right)^n + 1\right)^3}
\right) \nn
\hspace{-10mm}
&& \quad \left. + 36\left(4H\dot H + \ddot H\right)^2
\left(\frac{\frac{n(n-1)(n-2) c_1}{m^4} \left(R/m^2\right)^{n-3}}
{\left(c_2 \left(R/m^2\right)^n + 1\right)^2}
 - \frac{\frac{6 n^2(n-1) c_1 c_2 }{m^4} \left(R/m^2\right)^{2n-3} }
{\left(c_2 \left(R/m^2\right)^n + 1\right)^3}
+ \frac{\frac{6 n^3 c_1 c_2^2 }{m^4} \left(R/m^2\right)^{3n-3} }
{\left(c_2 \left(R/m^2\right)^n + 1\right)^4}\right) \right] \ .
\eea
Note that recently the observational bounds for $F(R)$ theories
have been discussed in Ref.~\cite{obs}.
We may also consider the $F({\cal G})$ gravity~\cite{Nojiri:2005jg}, whose
action is given by
\be
\label{FG1}
S=\int d^4 x\sqrt{-g}\left\{\frac{1}{2\kappa}\left(R + f_{\cal G}\left({\cal G}\right)\right)
+ {\cal L}_m\right\}\ ,
\ee
where ${\cal G}$ is the Gauss-Bonnet invariant:
\be
{\cal G}=R^2 -4 R_{\mu\nu} R^{\mu\nu} + R_{\mu\nu\xi\sigma}R^{\mu\nu\xi\sigma}\ .
\label{GB}
\ee
In the model, the effective energy density and pressure are given by
\bea
\label{EoS3}
\rho_{\rm eff} &=& \frac{1}{2\kappa^2}\left[{\cal G} f_{\cal G}'({\cal G})
 - f_{\cal G}({\cal G}) - 24^2 H^4 \left(2 {\dot H}^2 + H\ddot H + 4H^2 \dot H\right) f_{\cal G}'' \right]
+ \rho_{\rm matter}\ ,\nn
p_{\rm eff} &=& \frac{1}{2\kappa^2}\left[f_{\cal G}({\cal G}) + 24^2 H^2 \left( 3H^4 + 20 H^2 {\dot H}^2
+ 6 {\dot H}^3 + 4H^3 \ddot H + H^2 \dddot H \right) f_{\cal G}''({\cal G}) \right. \nn
&& \left. - 24^3 H^5 \left(2{\dot H}^2 + H \ddot H + 4 H^2 \dot H \right)^2 f_{\cal G}'''({\cal G})\right]
+ p_{\rm matter}\ .
\eea
In the FRW background, we find ${\cal G}=24\left(H^2 \dot H + H^4\right)$.
If we assume that the matter has a constant EoS parameter $w$, again,
Eq.~(\ref{EoS1}) has the following form:
\bea
\label{EoS4}
p_{\rm eff} &=& w \rho_{\rm eff} + G_{\cal G}\left(H, \dot H, \ddot H, \cdots\right)\ ,\nn
G_{\cal G}\left(H, \dot H, \cdots\right) &=&
\frac{1}{2\kappa^2}\left[(1+w) f_{\cal G}({\cal G}) - w{\cal G} f_{\cal G}'({\cal G}) \right. \nn
&& \left. + 24^2 H^2 \left( 3H^4 + 20 H^2 {\dot H}^2
+ 6 {\dot H}^3 + 4H^3 \ddot H + H^2 \dddot H
+ w H^4 \left(2 {\dot H}^2 + H\ddot H + 4H^2 \dot H\right) \right) f_{\cal G}''({\cal G}) \right. \nn
&& \left. - 24^3 H^5 \left(2{\dot H}^2 + H \ddot H + 4 H^2 \dot H \right)^2 f_{\cal G}'''({\cal G})\right]
\ .
\eea
An example is given by~\cite{Nojiri:2005jg}
\be
\label{GB9}
f_{\cal G}({\cal G})=f_0\left|{\cal G}\right|^{1/2}\ ,
\ee
where $f_0$ is a constant. In this case, one gets
\bea
\label{EoSAA1}
G_{\cal G}\left(H, \dot H, \cdots\right) &=&
\frac{f_0}{2\kappa^2}\left[\left(1+\frac{w}{2}\right) \left|{\cal G}\right|^{1/2} \right. \nn
&& \left. - 144 H^2 \left( 3H^4 + 20 H^2 {\dot H}^2
+ 6 {\dot H}^3 + 4H^3 \ddot H + H^2 \dddot H
+ w H^4 \left(2 {\dot H}^2 + H\ddot H + 4H^2 \dot H\right) \right)
\frac{1}{\left|{\cal G}\right|^{3/2}} \right. \nn
&& \left. + 9\cdot 24^2 H^5 \left(2{\dot H}^2 + H \ddot H + 4 H^2 \dot H \right)^2
\frac{{\cal G}}{2\left|{\cal G}\right|^{7/2}}\right]\ .
\eea
In the same way, one can obtain the function of modified gravity $G$ for
other models including non-local gravity~\cite{nonlocal}. It is interesting
to note that the perturbations of the above theory should be considered
by using the analogy with effective field theory~\cite{weinberg}.

We now study the general reconstruction program in terms of $G$ in
(\ref{EoS1}).
For simplicity, we assume that the matter has a constant EoS parameter $w$.
Using (\ref{GBR18B}), we find
\be
\label{EoS5}
G\left(H, \dot H, \cdots\right)  = - \frac{1}{\kappa^2}\left(2\dot H + 3(1+w)H^2 \right)\ .
\ee
Let the cosmology be given by $H=H(t)$. The right-hand side (r.h.s.)
of (\ref{EoS5}) is given by a function of $t$.
If the combination of $H$, $\dot H$, $\ddot H$ $\cdots$ in
$G\left(H, \dot H, \cdots\right)$ reproduces such a function, this kind of
cosmology could be realized.
As an illustrative example, we may consider the case in which $H(t)$ is
given by
\be
\label{EoS6}
H=h_0 + \frac{h_1}{t}\ ,
\ee
which gives
\be
\label{EoS7}
\dot H = - \frac{h_1}{t^2}\ ,\quad \ddot H = \frac{2h_1}{t^3}\ ,\quad
\cdots\ ,
\ee
and the r.h.s. in (\ref{EoS5}) is given by
\be
\label{EoS8}
- \frac{1}{\kappa^2}\left(2\dot H + 3(1+w)H^2 \right)
= - \frac{1}{\kappa^2}\left( 3(1+w)h_0^2 + \frac{6(1+w)h_0 h_1}{t}
+ \frac{-2h_1 + 3(1+w) h_1^2}{t^2} \right)\ .
\ee
For (trivial) example, if $G$ is given by
\be
\label{EoS9}
G\left(H,\dot H\right) = \frac{1}{\kappa^2}
\left\{
- 3(1+w)h_0^2 + 6(1+w)h_0 H
+ \left[ 2 - 3\left(1+w\right) h_0 \right] \dot H
\right\}\ ,
\ee
(\ref{EoS6}) is a solution.
Of course, there is large freedom of the choice of
$G\left(H,\dot H, \ddot H, \cdots\right)$,
but the form could be determined by the kind of the modified gravitational
theory which we are considering (for a detailed study of reconstruction for
various modified gravities, see~\cite{Nojiri:2006be, NOr}).
The important point is that the realistic history of the expansion of the
universe can be realized from the modified gravity reconstructed.

\section{Finite-time future singularities in $F(R)$ gravity \label{Sec3}}

In this section, we investigate $F(R)$-gravity models and show that some
models generate several known types of the finite-time future singularities.
This phenomenon is quite natural as modified gravity represented
as the Einstein gravity with the effective ideal fluid
with the phantom or quintessence-like EoS (see the explicit transformation
in Ref.~\cite{salv}).
In some cases, it is known that such ideal (quintessence or phantom) fluid
induces the finite-time future singularity.
We present several examples which predict the accelerating FRW solutions
ending at the finite-time future singularity by using the reconstruction
technique.
We demonstrate that not only the Big Rip but other three types of the
finite-time future singularities may appear.

As the first example, we consider the case of the Big Rip
singularity~\cite{innes}, where $H$ behaves as
\be
\label{frlv10}
H=\frac{h_0}{t_0 - t}\ ,
\ee
where $h_0$ and $t_0$ are positive constants and $H$ diverges at $t=t_0$.
To find the $F(R)$ gravity generating the Big Rip singularity, we use the
method of the reconstruction, namely, we construct $F(R)$ model realizing
{\it any} given cosmology using the technique proposed in Ref.~\cite{NOr}
(for the related study of reconstruction in $F(R)$ gravity, see~\cite{cortes}).
The action of $F(R)$ gravity with general matter is given as follows:
\be
\label{JGRG61}
S = \int d^4 x \sqrt{-g}\left\{F(R) + {\cal L}_{\rm matter}\right\} \ .
\ee
The action (\ref{JGRG61}) can be rewritten by using proper functions $P(\phi)$
and
$Q(\phi)$ of a scalar field $\phi$~\cite{NOr}:
\be
\label{JGRG62}
S=\int d^4 x \sqrt{-g} \left\{P(\phi) R + Q(\phi) + {\cal L}_{\rm matter}\right\}\ .
\ee
One may regard the scalar field $\phi$ as an auxiliary scalar field because
$\phi$ has no kinetic term. By the variation over $\phi$, we obtain
\be
\label{JGRG63}
0=P'(\phi)R + Q'(\phi)\ ,
\ee
which could be solved with respect to $\phi$ as $\phi=\phi(R)$.
By substituting $\phi=\phi(R)$ into the action (\ref{JGRG62}), we obtain the
action of $F(R)$ gravity given by
\be
\label{JGRG64}
F(R) = P(\phi(R)) R + Q(\phi(R))\ .
\ee

By assuming $\rho$, $p$ could be given by the corresponding sum of
matter with a constant EoS parameters $w_i$ and writing the scale factor
$a(t)$ as $a=a_0\e^{g(t)}$, where $a_0$ is a constant,
one gets the second rank differential equation:
\be
\label{JGRG68}
0 = 2 \frac{d^2 P(\phi)}{d\phi^2} - 2 g'(\phi) \frac{dP(\phi)}{d\phi} + 4g''(\phi) P(\phi)
+ \sum_i \left(1 + w_i\right) \rho_{i0} a_0^{-3(1+w_i)} \e^{-3(1+w_i)g(\phi)} \ .
\ee
If one can solve Eq.~(\ref{JGRG68}) with respect to $P(\phi)$, the form of
$Q(\phi)$ could be found as~\cite{NOr}
\be
\label{JGRG69}
Q(\phi) = -6 \left(g'(\phi)\right)^2 P(\phi) - 6g'(\phi) \frac{dP(\phi)}{d\phi}
+ \sum_i \rho_{i0} a_0^{-3(1+w_i)} \e^{-3(1+w_i)g(\phi)} \ .
\ee
Thus, it follows that any given history of the expansion of the universe can
be realized by some specific $F(R)$ gravity. Specific models unifying the
sequence: the early-time acceleration, radiation/matter-dominated stage and
dark energy epoch are constructed in
Refs.~\cite{NOr, Nojiri:2007cq, Nojiri:2007as}.

In case of (\ref{frlv10}), if we neglect the contribution from the matter,
the general solution of (\ref{JGRG68}) is given by
\be
\label{frlv11}
P(\phi) = P_+ \left(t_0 - \phi\right)^{\alpha_+}
+ P_- \left(t_0 - \phi\right)^{\alpha_-}\ ,\quad
\alpha_\pm \equiv \frac{- h_0 + 1 \pm \sqrt{h_0^2 - 10h_0 +1}}{2}\ ,
\ee
when $h_0 > 5 + 2\sqrt{6}$ or $h_0 < 5 - 2\sqrt{6}$ and
\be
\label{rlv12}
P(\phi) = \left(t_0 - \phi \right)^{-(h_0 + 1)/2}
\left( \hat A \cos \left( \left(t_0 - \phi \right) \ln \frac{ - h_0^2 + 10 h_0 -1}{2}\right)
+ \hat B \sin \left( \left(t_0 - \phi \right) \ln \frac{ - h_0^2 + 10 h_0 -1}{2}\right) \right)\ ,
\ee
when $5 + 2\sqrt{6}> h_0 > 5 - 2\sqrt{6}$.
Using (\ref{JGRG63}), (\ref{JGRG64}), and (\ref{JGRG69}), we find that
the form of $F(R)$ when $R$ is large is given by
\be
\label{rlv13}
F(R) \propto R^{1 - \alpha_-/2}\ ,
\ee
for $h_0 > 5 + 2\sqrt{6}$ or $h_0 < 5 - 2\sqrt{6}$ case and
\be
\label{rlv14}
F(R) \propto R^{\left(h_0 + 1\right)/4} \times \left(\mbox{oscillating parts}\right)\ ,
\ee
for $5 + 2\sqrt{6}> h_0 > 5 - 2\sqrt{6}$ case.

Let us investigate more general singularity~\cite{NO2008}
\be
\label{frlv9}
H \sim h_0 \left(t_0 - t\right)^{-\beta}\ .
\ee
Here, $h_0$ and $\beta$ are constants, $h_0$ is assumed to be positive and
$t<t_0$ because it should be for the expanding universe.
Even for non-integer $\beta<0$, some derivative of $H$ and therefore
the curvature becomes singular.
We should also note that Eq.(\ref{frlv9}) tells that the scale factor $a$ 
($H=\dot a/a$) behaves as 
\be
\label{scale}
a\sim \e^{ \frac{h_0 \left(t_0 - t\right)}{1 - \beta} + \cdots}\ .
\ee
Here, $\cdots$ expresses the regular terms. From (\ref{scale}), we find that 
if $\beta$ could not be any integer, the value of $a$, and therefore the value 
of the metric tensor,  would 
become complex number and include the imaginary part when 
$t>t_0$, which is unphysical. This could tell that the universe could end at 
$t=t_0$ even if $\beta$ could be negative or less than $-1$.

Since the case $\beta=1$ corresponds to
the Big Rip singularity, which has been investigated, we assume $\beta\neq 1$.
Furthermore, since $\beta=0$ corresponds to de Sitter space, which has no
singularity, we assume $\beta\neq 0$.
When $\beta>1$, the scalar curvature $R$ behaves as
\be
\label{rlv16B}
R \sim 12 H^2 \sim 12h_0^2 \left( t_0 - t \right)^{-2\beta}\ .
\ee
On the other hand, when $\beta<1$, the scalar curvature $R$ behaves as
\be
\label{rlv16C}
R \sim 6\dot H \sim 6h_0\beta \left( t_0 - t \right)^{-\beta-1}\ .
\ee
We may get the asymptotic solution for $P$ when $\phi\to t_0$.
\begin{itemize}
\item {\it $\beta>1$ case:} We find the following asymptotic expression of
$P(\phi)$:
\bea
\label{rlv29}
P(\phi) &\sim& \e^{\left(h_0/2\left(\beta - 1\right)\right)\left(t_0 - \phi\right)^{-\beta + 1}}
\left(t_0 - \phi\right)^{\beta/2}
\left(\tilde A \cos \left(\omega \left(t_0 - \phi\right)^{-\beta + 1}\right)
+ \tilde B \sin \left(\omega \left(t_0 - \phi\right)^{-\beta + 1}\right) \right)\ ,\nn
\omega &\equiv& \frac{h_0}{2\left(\beta - 1\right)}\ .
\eea
When $\phi\to t_0$, $P(\phi)$ tends to vanish.
Using (\ref{JGRG63}), (\ref{JGRG64}) and (\ref{JGRG69}),
we find that (at large $R$) $F(R)$ is given by
\be
\label{rlv29B}
F(R) \propto \e^{\left(h_0/2\left(\beta - 1\right)\right)
\left(\frac{R}{12h_0}\right)^{(\beta - 1)/2\beta}}
R^{-1/4}\times\left( \mbox{oscillating part} \right)\ .
\ee
\item {\it $1 > \beta > 0$ case:} We find the following asymptotic
expression of $P(\phi)$:
\be
\label{rlv31}
P(\phi) \sim B \e^{-\left(h_0/2\left(1 - \beta\right)\right)
\left(t_0 - \phi\right)^{1-\beta}}\left(t_0 - \phi\right)^{\left(\beta + 1\right)/8}\ .
\ee
Hence $F(R)$ is given by
\be
\label{rlv31C}
F(R) \sim \e^{-\left(h_0/2\left(1-\beta\right)\right)
\left( - 6\beta h_0 R \right)^{(\beta - 1)/(\beta + 1)} } R^{7/8}\ .
\ee
\item {\it $\beta<0$ case:} The asymptotic expression of $P(\phi)$ is as
follows:
\be
\label{rlv33}
P(\phi) \sim A \e^{-\left(h_0/2\left(1 - \beta\right)\right)
\left(t_0 - \phi\right)^{1-\beta}}
\left(t_0 - \phi\right)^{- \left(\beta^2 - 6\beta + 1\right)/8}\ .
\ee
Thus $F(R)$ is given by
\be
\label{rlv35}
F(R) \sim
\left( -6h_0 \beta R \right)^{\left(\beta^2 + 2\beta + 9\right)/8\left(\beta +1\right)}
\e^{-\left(h_0/2\left(1 - \beta\right)\right)
\left( -6h_0 \beta R \right)^{\left(\beta-1\right)/\left(\beta + 1\right)}} \ .
\ee
Note that $-6h_0 \beta R >0$ when $h_0, R>0$.
\end{itemize}

We found the behavior of the scalar curvature $R$ from that of $H$ in
(\ref{frlv9}), namely, when $\beta>1$, $R$ behaves as in (\ref{rlv16B}),
and when $\beta<1$, $R$ behaves as in (\ref{rlv16C}).
Now, conversely we consider the behavior of $H$ from that of $R$.
When $R$ behaves as
\be
\label{R1}
R \sim 6\dot H \sim R_0 \left( t_0 - t \right)^{-\gamma}\ ,
\ee
if $\gamma>2$, which corresponds to $\beta = \gamma/2 >1$,
$H$ behaves as
\be
\label{R2}
H \sim \sqrt{\frac{R_0}{12}} \left( t_0 - t \right)^{-\gamma/2}\ ,
\ee
if $2>\gamma>1$, which corresponds to $1> \beta = \gamma -1 >0$,
$H$ is given by
\be
\label{R3}
H \sim \frac{R_0}{6\left( \gamma - 1\right)} \left( t_0 - t \right)^{-\gamma + 1}\ ,
\ee
and if $\gamma<1$, which corresponds to $\beta = \gamma -1 <0$, one
obtains
\be
\label{R4}
H \sim H_0 + \frac{R_0}{6\left( \gamma - 1\right)} \left( t_0 - t \right)^{-\gamma + 1}\ .
\ee
Here, $H_0$ is an arbitrary constant, and it does not affect
the behavior of $R$.
$H_0$ is chosen to vanish in (\ref{frlv9}).
>From $H=\dot a(t)/a(t)$, if $\gamma>2$, we find
\be
\label{R5}
a(t) \propto \exp \left( \left(\frac{2}{\gamma} -1 \right)
\sqrt{\frac{R_0}{12}} \left( t_0 - t \right)^{-\gamma/2 + 1}\right)\ ,
\ee
when $2>\gamma>1$, $a(t)$ behaves as
\be
\label{R6}
a(t) \propto \exp \left( \frac{R_0}{6\gamma\left( \gamma - 1\right)}
\left( t_0 - t \right)^{-\gamma}\right)\ ,
\ee
and if $\gamma<1$,
\be
\label{R7}
a(t) \propto \exp \left( H_0 t + \frac{R_0}{6\gamma\left( \gamma - 1\right)}
\left( t_0 - t \right)^{-\gamma}\right)\ .
\ee
In any case, there appears a sudden future singularity~\cite{sudden} at
$t=t_0$.

Since the second term in (\ref{R4}) is smaller than the first one, one
may solve (\ref{JGRG68}) asymptotically as follows:
\be
\label{RR1}
P\sim P_0 \left( 1 + \frac{2h_0}{1-\beta}\left(t_0 - \phi\right)^{1-\beta}\right)\ ,
\ee
with a constant $P_0$, which gives
\be
\label{RR2}
F(R) \sim F_0 R + F_1 R^{2\beta/\left(\beta + 1\right)}\ .
\ee
Here, $F_0$ and $F_1$ are constants.
When $0>\beta>-1$, we find $2\beta/\left(\beta + 1\right)<0$.
On the other hand, when $\beta<-1$, we obtain
$2\beta/\left(\beta + 1\right)>2$.
As we saw in (\ref{rlv13}), for $\beta<-1$, $H$ diverges when $t\to t_0$.
Since we reconstruct $F(R)$ so that the behavior of $H$ could be recovered,
the $F(R)$ generates the Big Rip singularity when $R$ is large.
Thus, even if $R$ is small, the $F(R)$ generates a singularity
where higher derivatives of $H$ diverge.

Even for $F(R)$ gravity, we may define the effective energy density
$\rho_{\rm eff}$ and pressure $p_{\rm eff}$ as (\ref{GBR18B}).
We now assume $H$ behaves as (\ref{frlv9}).
For $\beta> 1$, when $t\to t_0$,
$a\sim \exp( h_0\left( t_0 - t \right)^{1-\beta}/\left( \beta -1 \right) )
\to \infty$ and
$\rho_{\rm eff} ,\, |p_{\rm eff}| \to \infty$.
If $\beta=1$, we find
$a\sim \left(t_0 - t\right)^{-h_0} \to \infty$ and
$\rho_{\rm eff} ,\, |p_{\rm eff}| \to \infty$.
If $0<\beta<1$, $a$ goes to a constant but $\rho ,\, |p| \to \infty$.
If $-1<\beta<0$, $a$ and $\rho$ vanish but $|p_{\rm eff}| \to \infty$.
When $\beta<0$, instead of (\ref{frlv9}), as in (\ref{R3}), one may assume
\be
\label{R13}
H \sim H_0 + h_0 \left(t_0 - t\right)^{-\beta}\ .
\ee
Hence, if $-1<\beta<0$, $\rho_{\rm eff}$ has a finite value $3H_0^2/\kappa^2$
in the limit $t\to t_0$.
If $\beta<-1$ but $\beta$ is not any integer, $a$ is finite and
$\rho_{\rm eff}$ and $p_{\rm eff}$ vanish if $H_0=0$ or $\rho_{\rm eff}$
and $p_{\rm eff}$ are finite if $H_0\neq 0$ but higher derivatives of $H$
diverge.
We should note that the leading behavior of the scalar curvature $R$ does not
depend on $H_0$ in (\ref{R13}), and that the second term in (\ref{R13}) is
relevant to the leading behavior of $R$. We should note, however, that $H_0$
is relevant to the leading behavior of the effective energy density
$\rho_{\rm eff}$ and the scale factor $a$.

In Ref.~\cite{Nojiri:2005sx}, the classification of the finite-time future
singularities has been suggested in the following way:
\begin{itemize}
\item Type I (``Big Rip'') : For $t \to t_s$, $a \to \infty$,
$\rho \to \infty$ and $|p| \to \infty$. This also includes the case of
$\rho$, $p$ being finite at $t_s$.
\item Type II (``sudden'') : For $t \to t_s$, $a \to a_s$,
$\rho \to \rho_s$ and $|p| \to \infty$
\item Type III : For $t \to t_s$, $a \to a_s$,
$\rho \to \infty$ and $|p| \to \infty$
\item Type IV : For $t \to t_s$, $a \to a_s$,
$\rho \to 0$, $|p| \to 0$ and higher derivatives of $H$ diverge.
This also includes the case in which $p$ ($\rho$) or both of $p$ and $\rho$
tend to some finite values, while higher derivatives of $H$ diverge.
\end{itemize}
Here, $t_s$, $a_s (\neq 0)$ and $\rho_s$ are constants.
We now identify $t_s$ with $t_0$.
The Type I corresponds to $\beta>1$ or $\beta=1$ case, Type II to $-1<\beta<0$
case, Type III
to $0<\beta<1$ case, and Type IV to $\beta<-1$ but $\beta$ is not any
integer number.
When the phantom dark energy was studied, it was found that the Big Rip
type (type I) singularity could occur. After that, it was pointed out that
there could be another kind of singularity corresponding to
type II~\cite{sudden}. Then, it was found in Ref.~\cite{Nojiri:2005sx}
that there are other kinds of singularities corresponding to
type III and IV. Note that if only higher derivatives of Hubble rate
diverge,
then some combination of curvature invariants also diverges what leads to
singularity.
Thus, we have constructed several examples of $F(R)$ gravity showing any type
of the above finite-time future singularities. This is natural because it is
known that modified gravity may lead to the effective phantom/quintessence
phase~\cite{review}, while the phantom/quintessence-dominated universe may
end up with the finite-time future singularity.

The reconstruction method also tells that there appears Type I singularity
for $F(R) = R + \tilde{\alpha} R^n$ with $n>2$ and Type III singularity for
$F(R) = R - \tilde{\beta} R^{-n}$ with $n>0$, where $\tilde{\alpha}$ and
$\tilde{\beta}$ are constants.
Note, however, that even if some specific model contains the finite-time
future singularity, one can always make the reconstruction of the model
in the remote past in such a way that the finite-time future singularity
disappears. Usually, positive powers of the curvature (polynomial structure)
help to make the effective quintessence/phantom phase to become transient and
to avoid the finite-time future singularities.
The corresponding examples are presented in Refs.~\cite{NO2008, abdalla}.

\section{Absence of singularity in modified gravity \label{Sec4}}

In this section, we study various scenarios to resolve the finite-time future
singularities. The most natural scenario is based on additional
modification of the inhomogeneous EoS or the gravitational action by the term
which is not relevant currently.
In fact, however, such a term which may be relevant
at very early or very late universe may resolve the singularity.
We note that the presence of such a next-to-leading order term does not
conflict with the known local tests. Another scenario is related with
the account of quantum effects becoming relevant near to singularity.

Let us consider the conditions for $G$ in (\ref{EoS1}), which prevents the
finite-time future singularities appearing. For simplicity, we only consider
the case in which the matter has a constant EoS parameter $w$.
A key equation is Eq.~(\ref{EoS5}).
If Eq.~(\ref{EoS5}) becomes inconsistent for any singularities,
the singularity could not be realized.
First, we put $G=0$ in (\ref{EoS5}).
In this case, a solution satisfying (\ref{EoS5}) is given by
\be
\label{E1}
H= \frac{- \frac{2}{3(1+w)}}{t_0 - t}\ .
\ee
If $w<-1$, i.e., the phantom phase, there appears a Big Rip singularity.
We may now assume $w>-1$.
Even if $w$ is not constant, in case $-\infty<w<-1$, there occurs the Big Rip 
type singularity, where $H$ diverges in finite future. For example, we may 
consider the case in which $w$ behaves as $w=-1 - w_0\left(t_0 - t\right)^\eta$ when $t\sim t_0$. Here, $w_0$ and $\eta$ are constants. In order that $w$ 
could be less than $-1$, $w_0$ should be positive. 
Hence the Hubble rate behaves as $H\sim \left(t_0 - t\right)^{-1 - \eta}$. 
Thus, if $\eta>-1$, $H$ diverges and the Big Rip type singularity could occur. 
We should note that when $\eta>0$, $w\to -1$ in the limit of $t\to t_0$.

If $H$ evolves as (\ref{R13}), the r.h.s. in (\ref{EoS5}) behaves as
\be
\label{E2}
- \frac{1}{\kappa^2}\left(2\dot H + 3(1+w)H^2 \right) \sim
\left\{
\begin{array}{lcl}
 -\frac{3(1+w)h_0^2}{\kappa^2}\left(t_0 - t\right)^{-2\beta} & \mbox{when} & \beta>1 \\
 -\frac{2\beta h_0 + 3(1+w)h_0^2}{\kappa^2}\left( t_0 - t \right)^{-2} && 0<\beta<1 \\
 -\frac{2\beta h_0}{\kappa^2} \left(t_0 - t\right)^{-\beta - 1} && 0>\beta>-1
\end{array} \right. \ .
\ee

If $\beta>-1$, $\beta\neq 0$, which correspond to Type I, II and III
singularities, the l.h.s. on (\ref{E2}), i.e.,
$G\left(H, \dot H, \cdots\right)$ in Eq.~(\ref{EoS5}), diverges.
One way to prevent such a singularity appearing could be that $G$ is bounded.
In this case, Eq.~(\ref{EoS5}) becomes inconsistent with the behavior of
the r.h.s. of (\ref{E2}).
An example is as follows:
\be
\label{E3}
G=G_0 \frac{1 + a H^2}{1+ b H^2}\ ,
\ee
where $a$ and $b$ are positive constants and $G_0$ is a constant.

We should also note that when $\beta>0$, which corresponds to Type I or III
singularity, the r.h.s. of (\ref{E2}) becomes negative. Hence, if $G$ is
positive for large $H$, the singularity could not be realized.

Another possibility is that if $G$ contains the term like
$\sqrt{ 1 - a^2 H^2}$, which becomes imaginary for large $H$, (\ref{E2}) could
become inconsistent.
Thus, singularities where the curvature blows up (Type I, II, III) could be
prevented appearing. This mechanism could be applied even if $w<-1$.
One can add an extra term as
\be
\label{Gl1}
G_1 (H) = G_0 \left(\sqrt{  1 - \frac{H^2}{H_0^2} } - 1\right)\ ,
\ee
to $G$. Here, $G_0$ and $H_0$ are constants. If we choose $H_0$ to be large
enough, $G_1$
is not relevant for the small curvature but relevant for large scale and
prohibits the curvature singularity appearing.
Moreover, it is easy to check the EoS of the effective ideal fluid induced by
modified gravity. Some indications of the presence/absence of
the singularities can be found here. Nevertheless, one should not forget
that the description of the effective ideal fluid corresponds to
the consideration of the theory in another (Einstein) frame.
The transformation of the point where the singularity appears from one frame
to another may be tricky in modified gravity~\cite{briscese}.

As the universe approaches to the singularity, the curvature may become large
again. Hence, it is necessary to take into account quantum effects (or even
those of modified gravity through effective action approach~\cite{percacci}).
One may include the massless quantum effects by taking account of the
contribution of the conformal anomaly as the back reaction near
the singularity.
The conformal anomaly $T_A$ has the following well-known
form~\cite{Nojiri:2005sx}:
\be
\label{OVII}
T_A=b\left(F_\mathrm{W} + \frac{2}{3}\Box R\right) + b' {\cal G}
+ b''\Box R\ .
\ee
Here, $F_\mathrm{W}$ is the square of 4d Weyl tensor, and it is given by
\bea
\label{GF}
F_\mathrm{W} &=& \frac{1}{3}R^2 -2 R_{\mu\nu}R^{\mu\nu}+
R_{\mu\nu\rho\sigma}R^{\mu\nu\rho\sigma}\ . \nn
\eea
In general, with $N$ scalar, $N_{1/2}$ spinor, $N_1$ vector fields,
$N_2$ ($=0$ or $1$)
gravitons and $N_{\rm HD}$ higher derivative conformal scalars, $b$ and
$b'$ are given by~\cite{Nojiri:2005sx}
\bea
\label{bs}
&& b= \frac{N +6N_{1/2}+12N_1 + 611 N_2 - 8N_{\rm HD}}{120(4\pi)^2}\ ,\nn
&& b'=- \frac{N+11N_{1/2}+62N_1 + 1411 N_2 -28 N_{\rm HD}}{360(4\pi)^2}\ .
\eea
As is seen, $b>0$ and $b'<0$ for the usual matter except the higher
derivative conformal scalars.
Notice that $b''$ can be shifted by the finite renormalization of the
local counter term $R^2$, so that $b''$ can be an arbitrary coefficient.
For the FRW universe, we find
\be
\label{CA2}
F_\mathrm{W} = 0\ ,\quad {\cal G} = 24\left(\dot H H^2 + H^4\right)\ .
\ee
In terms of the corresponding energy density $\rho_A$ and pressure $p_A$,
$T_A$ is given by~\cite{Nojiri:2005sx}
\be
\label{CA3}
T_A = - \rho_A + 3p_A \ ,
\ee
and the energy density could be conserved
\be
\label{CA4}
\dot\rho_A + 3H \left( \rho_A + p_A \right) = 0\ .
\ee
Hence we find
\be
\label{CA5}
\rho_A = - a^{-4}\int dt a^4 H T_A\ ,\quad
p_A = \frac{T_A}{3} - \frac{a^{-4}}{3}\int dt a^4 H T_A\ .
\ee
By using the above expressions, Eq.~(\ref{GBR18B}) could be modified as
\be
\label{GBR18BB}
\rho_{\rm eff} + \rho_A = \frac{3}{\kappa^2}H^2\ ,\quad
p_{\rm eff} + p_A = -\frac{1}{\kappa^2}\left(2\dot H + 3H^2\right)\ .
\ee
If we redefine the effective energy density and pressure as
\be
\label{CA6}
\tilde \rho_{\rm eff} = \rho_{\rm eff} + \rho_A \ ,\quad
\tilde p_{\rm eff} = p_{\rm eff} + p_A\ ,
\ee
Eq.~(\ref{EoS2}) could be modified as
\bea
\label{CA7}
\tilde p_{\rm eff} &=& w \tilde \rho_{\rm eff}
+ \tilde G\left(H, \dot H, \ddot H, \cdots\right)\ ,\nn
\tilde G &\equiv& G\left(H, \dot H, \cdots\right)
+ \frac{T_A}{3} - \left(\frac{1}{3} + w \right) a^{-4}
\int dt a^4 H T_A\ .
\eea
It has been indicated~\cite{quantum} that the explicit account of such quantum
effects tends to moderate the singularity, to make the space-time less
singular or at least to delay the Rip time. Clearly, these scenarios may be
applied to any specific modified gravity under consideration.
Furthermore, there remains the possibility that the action of modified gravity
is changed by the term which is not relevant now and does not influence the
local tests in such a theory currently. This kind of term, however, may be
relevant at the very early universe or at the very late universe in such a
way that the universe avoids approaching to the singularity (in other words,
say, the phantom phase becomes transient as in the model of
Ref.~\cite{abdalla}).

As the explicit realization of the above modification, one can consider the
effective ideal fluid with the complicated EoS depending on the Hubble rate:
\be
\label{WW1}
p_{\rm eff} = w\left(H,\dot H, \cdots \right)\rho_{\rm eff}\ ,
\ee
namely, the EoS parameter could be a function of $H$, $\dot H$, $\ddot H$,
$\cdots$. The specific form of the dependence on the Hubble rate is defined
by the equivalent model of modified gravity.
Using (\ref{GBR18B}), we find
\be
\label{WW2}
0=2\dot H + 3H^2 + 3w\left(H,\dot H, \cdots \right) H^2\ .
\ee
If we take $H$ to be a constant $H_0$, Eq.~(\ref{WW2}) is reduced to
\be
\label{WW3}
0=1 + w\left(H_0,\dot H=0, \cdots \right)\ ,
\ee
which is an algebraic equation and hence has a solution, e.g., there could be
a de Sitter space solution.
For example, if $w$ is given by
\be
\label{WW4}
w\left(H,\dot H, \cdots \right) = \frac{H^2}{h_0^2} + f\left(\dot H\right)\ ,
\ee
it follows
\be
\label{WW5}
H=h_0 \sqrt{1 - f(0)}\ .
\ee
As a special case, if $w$ is given by
\be
\label{WW6}
w\left(H,\dot H\right) = -1 + \frac{2\dot H}{3H^2}\ ,
\ee
Eq.~(\ref{WW2}) is trivially satisfied for any $H$, and therefore any
cosmology can be a solution.
This theory, however, has no predictive power.
One non-trivial example is as follows:
\be
\label{WW7}
w(H)= - 1 + \frac{2\beta}{3h_0^{1/\beta}} H^{-1 + 1/\beta}\ .
\ee
The solution of (\ref{WW2}) is given by the exact form of
(\ref{frlv9}), i.e., $H=h_0 \left(t_0 - t\right)^{-\beta}$. Hence, any type of
the finite-time future singularities can be realized by the form of $w$
in (\ref{WW6}).
Another non-trivial example is expressed by the following form:
\be
\label{WW8}
w(H) = -1 - \frac{2\left(h_i - h_l\right)}{3t_0 H^2}
\left\{ 1 - \left(\frac{h_i + h_l}{h_i - h_l} - \frac{2H}{h_i - h_l}\right)^2\right\}\ ,
\ee
where $t_0$, $h_i$ and $h_l$ are constants satisfying $h_i\gg h_l > 0$.
An exact solution of (\ref{WW2}) is given by
\be
\label{WW9}
H=\frac{h_i + h_l}{2} - \frac{h_i - h_l}{2}\tanh \frac{t}{t_0}\ .
\ee
In the limit $t\to -\infty$, we find $H\to h_i$. On the other hand, in the
limit $t\to \infty$, $H\to h_l$. Hence the de Sitter universe could be
realized in both limits $t\to \pm\infty$. Thus,
we may identify the limit of $t\to -\infty$ with inflation and the limit of
$t\to \infty$ with the late time acceleration.
Adding the effective ideal fluid with such a EoS to the model admitting
the singularity, one can always check if this addition resolves it.

\section{Finite-time future singularities in scalar-Gauss-Bonnet and modified
Gauss-Bonnet gravities \label{Sec5}}

In this section, we discuss the scalar-Gauss-Bonnet and modified
Gauss-Bonnet gravities which predict the late-time acceleration ending at a
finite-time future singularity.
We present the corresponding effective potentials by
using the reconstruction technique.

Let us consider the
reconstruction~\cite{Nojiri:2006be, Nojiri:2006je, Cognola:2006sp}
of the string-inspired scalar-Gauss-Bonnet gravity proposed as dark
energy in Refs.~\cite{Nojiri:2005vv, GB} for the investigation of the
finite-time future singularities:
\be
\label{GBR1}
S=\int d^4 x \sqrt{-g}\left[ \frac{R}{2\kappa^2} - \frac{1}{2}\partial_\mu \phi \partial^\mu \phi
 - V(\phi) - \xi(\phi) {\cal G}\right]\ .
\ee

The theory depends on two scalar potentials.
The explicit example motivated by string considerations is the
following~\cite{Nojiri:2005vv}:
\be
\label{MOS}
V(\phi) = V_0 \e^{-2\phi/\phi_0}\ ,\quad \xi(\phi)=\xi_0 \e^{2\phi/\phi_0}\ ,
\ee
where $V_0$, $\phi_0$ and $\xi_0$ are constant parameters.
By choosing the parameters properly,
dark energy universe ending at the Big Rip singularity emerges.

There exists another example corresponding to the following choice of the
potentials~\cite{Cognola:2006sp}:
\bea
\label{LGB3}
V(\phi) &=& \frac{3}{\kappa^2}\left( g_0 + \frac{g_1}{t_0}
\e^{- \phi/\phi_0}\right)^2 - \frac{g_1}{\kappa^2t_0^2} \e^{- 2\phi/\phi_0}
- 3U_0\left(g_0 + \frac{g_1}{t_0}\e^{- \phi/\phi_0}\right)\e^{g_1 \phi/\phi_0}
\e^{g_0 t_0\e^{\phi/\phi_0}}\ ,\nn
\xi (\phi) &=& \frac{U_0}{8}\int^{t_0 \e^{\phi/\phi_0}} dt_1 \left( g_0 +
\frac{g_1}{t_1}\right)^{-2}
\left(\frac{t}{t_0}\right)\e^{g_0 t}\ .
\eea
Here $g_0$, $g_1$, and $U_0$ are constants.
The Hubble rate is given by
\be
\label{LGB4}
H=g_0 + \frac{g_1}{t}\ ,\quad \phi=\phi_0 \ln \frac{t}{t_0}\ .
\ee
Hence, when $t$ is small, the second term in the expression of $H$ in
(\ref{LGB4}) dominates and the scale factor behaves as $a\sim t^{g_1}$.
Thus, if $g_1=2/3$, a matter-dominated period, in which a scalar may be
identified with matter, could be realized.
On the other hand, when $t$ is large, the first term of $H$ in (\ref{LGB4})
dominates and the Hubble rate $H$ becomes constant.
Consequently, the universe is asymptotically de Sitter space, which is an
accelerating universe.

In the case without matter, the equations of motion in the FRW metric are
given by
\bea
\label{GBR3}
0&=& - \frac{3}{\kappa^2}H^2 + \frac{1}{2}{\dot\phi}^2 + V(\phi) + 24 H^3 \frac{d \xi(\phi(t))}{dt}\ ,\\
\label{GBany5}
0&=& \frac{1}{\kappa^2}\left(2\dot H + 3 H^2 \right) + \frac{1}{2}{\dot\phi}^2 - V(\phi)
 - 8H^2 \frac{d^2 \xi(\phi(t))}{dt^2} - 16H \dot H \frac{d\xi(\phi(t))}{dt}
 - 16 H^3 \frac{d \xi(\phi(t))}{dt}\ ,\\
0&=&\ddot \phi + 3H\dot \phi + V'(\phi) + \xi'(\phi) {\cal G}\ .
\eea
One may rewrite the above equations to the following form:
\bea
\label{GBR4}
0&=&\frac{2}{\kappa^2}\dot H + {\dot\phi}^2 - 8H^2 \frac{d^2 \xi(\phi(t))}{dt^2}
 - 16 H\dot H \frac{d\xi(\phi(t))}{dt} + 8H^3 \frac{d\xi(\phi(t))}{dt} \nn
&=&\frac{2}{\kappa^2}\dot H + {\dot\phi}^2
 - 8a\frac{d}{dt}\left(\frac{H^2}{a}\frac{d\xi(\phi(t))}{dt}\right)\ .
\eea
The potentials are expressed as
\bea
\label{GBR5}
V(\phi(t)) &=& \frac{3}{\kappa^2}H(t)^2 - \frac{1}{2}{\dot\phi (t)}^2
 - 3a(t) H(t) \int^t \frac{dt_1}{a(t_1)}
\left(\frac{2}{\kappa^2}\dot H (t_1) + {\dot\phi(t_1)}^2 \right)\ , \nn
\xi(\phi(t))&=&\frac{1}{8}\int^t dt_1 \frac{a(t_1)}{H(t_1)^2}
\int^{t_1} \frac{dt_2}{a(t_2)}
\left(\frac{2}{\kappa^2}\dot H (t_2) + {\dot\phi(t_2)}^2 \right)\ .
\eea
Equations~(\ref{GBR5}) show that if we consider the theory including two
functions $g(t)$ and $f(\phi)$ as
\bea
\label{GBR6}
V(\phi) &=& \frac{3}{\kappa^2}g'\left(f(\phi)\right)^2 - \frac{1}{2f'(\phi)^2}
 - 3g'\left(f(\phi)\right) \e^{g\left(f(\phi)\right)}  U(\phi) \ ,\nn
\xi(\phi) &=& \frac{1}{8}\int^\phi d\phi_1 \frac{f'(\phi_1)
\e^{g\left(f(\phi_1)\right)} }{g'\left(f(\phi_1)\right)^2} U(\phi_1)\ ,\nn
U(\phi) &\equiv& \int^\phi d\phi_1 f'( \phi_1 )
\e^{-g\left(f(\phi_1)\right)} \left(\frac{2}{\kappa^2}
g''\left(f(\phi_1)\right) + \frac{1}{f'(\phi_1 )^2} \right)\ ,
\eea
the solution of the field equations is given by
\be
\label{GBR7}
\phi=f^{-1}(t)\quad \left(t=f(\phi)\right)\ ,\quad a=a_0\e^{g(t)}\ \left(H= g'(t)\right)\ .
\ee
It is easy to include matter with constant $w=w_m$. In this case, it is
enough to consider the theory where $U(\phi)$ in $V(\phi)$ and $\xi(\phi)$
in (\ref{GBR6}) is replaced by
\be
\label{GBR8}
U(\phi) = \int^\phi d\phi_1 f'(\phi_1)
\e^{-g\left(f(\phi_1)\right)} \left(\frac{2}{\kappa^2}g''\left(f(\phi_1)\right) + \frac{1}{f'(\phi_1 )^2}
+ (1+w_m)g_0\e^{-3(1+w_m)g\left(f(\phi_1)\right)}\right) \ .
\ee
We obtain the previous solution with
$a_0=\left(g_0/\rho_0\right)^{-1/3(1+w_m)}$.

As a cousin of the scalar-Gauss-Bonnet theory, we may consider
the modified Gauss-Bonnet
theory~\cite{Nojiri:2005jg, Nojiri:2005am, Cognola:2006eg},
whose action is given by
\be
\label{GBR9}
S=\int d^4 x \sqrt{-g}\left[ \frac{R}{2\kappa^2} + F({\cal G}) \right]\ .
\ee
By introducing an auxiliary scalar field $\phi$, we can rewrite the action
(\ref{GBR9}) to a form similar to the action of the scalar Gauss-Bonnet theory
in (\ref{GBR1}):
\be
\label{GBR10}
S=\int d^4 x \sqrt{-g}\left[ \frac{R}{2\kappa^2} - V(\phi) - \xi(\phi) {\cal G}\right]\ .
\ee
In fact, by solving the equation for $\phi$:
\be
\label{GBR11}
0=V'(\phi) + \xi'(\phi) {\cal G} \ ,
\ee
with respect to $\phi$ as $\phi= \phi({\cal G})$ and substituting the
expression of $\phi$ into the action (\ref{GBR10}), we re-obtain
the action~(\ref{GBR9}) where $F({\cal G})$ is given by
\be
\label{GBR12}
F({\cal G}) \equiv - V\left(\phi({\cal G})\right) + \xi\left(\phi({\cal G})\right){\cal G}\ .
\ee

An example~\cite{Nojiri:2005jg} is given by (\ref{GB9}).
When $F_0^2>3/2\kappa^4$, there are two solutions, which describe effective
phantom universe
and admit the Big Rip singularity.
When $F_0^2<3/2\kappa^4$, we have a solution, which describes
the effective quintessence, and the other solution, which does effective
phantom.

Another example with dust, whose energy density behaves as
$\rho = \rho_{0d} a^{-3}$, is as follows~\cite{Cognola:2006sp}:
\bea
\label{ex2}
V(\phi) &=& \frac{2C^2}{\kappa^2}\coth^2 (C\phi) - 3CU_0 \coth (C\phi) \sinh^{2/3} (C\phi)\ ,\nn
\xi(\phi)  &=& \frac{U_0}{8}\int^\phi d\phi_1 \sinh^{-4/3} (C\phi) \cosh^2 (C\phi)\ .
\eea
Here, $C$ and $U_0$ are constants. The explicit solution is given by
\be
\label{ex1}
a(t)=a_0\e^{g(t)}\ ,\quad g(t)=\frac{2}{3}\ln\left(\sinh\left(C t\right)\right)\ ,\quad
\rho_{0d} = \frac{27a_0^{3} C}{4\kappa^2}\ .
\ee
Eq.~(\ref{ex1}) indicates that, when $\phi=t$ is small, $g(\phi)$ behaves as
$g(\phi)\sim (2/3)\ln \phi$
and, therefore, the Hubble rate behaves as
$H(t)=g'(t)\sim (2/3)/t$, which surely reproduces the matter-dominated
phase. On the other hand, when $\phi=t$ is large,
$g(\phi)$ behaves as $g\sim (2/3) (C\phi)$, namely, $H \sim 2C/3$ and
the universe asymptotically goes to de Sitter space.
Hence, the model given by (\ref{ex2}) {\it with matter} shows
the transition from the matter-dominated phase to the accelerating universe,
which is asymptotically de Sitter space.

We may consider the reconstruction of $F({\cal G})$
gravity~\cite{Nojiri:2006be, Cognola:2006sp}.
The field equations in the FRW background are given by
\bea
\label{GBR13}
0&=& - \frac{3}{\kappa^2}H^2 + V(\phi) + 24 H^3 \frac{d \xi(\phi(t))}{dt}\ ,\\
\label{GBR14}
0&=& \frac{1}{\kappa^2}\left(2\dot H + 3 H^2 \right) - V(\phi)
 - 8H^2 \frac{d^2 \xi(\phi(t))}{dt^2}
 - 16H \dot H \frac{d\xi(\phi(t))}{dt} - 16 H^3 \frac{d \xi(\phi(t))}{dt}\ ,
\eea
which could be rewritten as
\be
\label{GBR15}
\xi(\phi(t)) = \frac{1}{8}\int^t dt_1 \frac{a(t_1)}{H(t_1)^2} W(t_1) \ ,\quad
V(\phi(t)) = \frac{3}{\kappa^2}H(t)^2 - 3a(t) H(t) W(t) \ , \quad
W(t) \equiv \frac{2}{\kappa^2} \int^{t} \frac{dt_1}{a(t_1)} \dot H (t_1) \ .
\ee
Since there is no kinetic term of $\phi$, there is a freedom to redefine $\phi$
as $\phi\to \varphi=\varphi(\phi)$. By using the redefinition, we may choose
the scalar field $\phi$ as a time coordinate: $\phi=t$.
Thus one gets
\be
\label{GBR16}
V(\phi) = \frac{3}{\kappa^2}g'\left(\phi\right)^2
 - 3g'\left(\phi\right) \e^{g\left(\phi\right)} U(\phi) \ , \quad
\xi(\phi) = \frac{1}{8}\int^\phi d\phi_1 \frac{\e^{g\left(\phi_1\right)} }{g'(\phi_1)^2}
U(\phi_1)\ ,\quad
U(\phi) \equiv \frac{2}{\kappa^2}\int^\phi d\phi_1 \e^{-g\left(\phi_1\right)} g''\left(\phi_1\right) \ .
\ee
The solution is given by
\be
\label{GBR16b}
a=a_0\e^{g(t)}\ \left(H= g'(t)\right)\ .
\ee
As in the case of the scalar-Gauss-Bonnet theory, we may include
matter.

Let us apply the reconstruction program to the study of the finite-time future
singularities, where the Hubble rate behaves as
\be
\label{GBR17}
H \sim h_0 \left(t_s - t\right)^{-\beta}\ .
\ee
First, we consider the modified Gauss-Bonnet gravity.
For $\beta=1$ and $h_0>0$ case, namely, usual Big Rip singularity, since
$g'(\phi)= H(\phi)$, we find
\be
\label{GBR19}
g(\phi) = - h_0 \ln \frac{t_s - \phi}{t_0}\ ,
\ee
where $t_0$ is a constant of the integration. $U(\phi)$ in (\ref{GBR16}) has
the following form:
\be
\label{GBR20}
U(\phi) = \frac{2}{\kappa^2}\int^\phi d\phi_1 \left(\frac{t_s - \phi_1}{t_0}\right)^{h_0}
\frac{h_0}{\left(t_s - \phi_1\right)^2}
= \left\{
\begin{array}{ll}
U_0 - \frac{2h_0}{\kappa^2{t_0}^{h_0}\left(h_0 -1\right)} \left(t_s - \phi\right)^{h_0-1}
\ & \mbox{when}\ h_0 \neq 1 \\
\frac{2}{\kappa^2 t_0} \ln \frac{t_s - \phi}{t_1}
\ & \mbox{when}\ h_0 = 1
\end{array}\right. \ .
\ee
Here, $U_0$ and $t_1$ are constants of the integration.
At the next step, one gets
\bea
\label{GBR21}
V(\Phi) &=& \left\{ \begin{array}{ll}
 - \frac{3h_0 t_0^{h_0} U_0}{\Phi^{h_0 + 1}}
+ \frac{3h_0^2 \left(h_0 + 1\right)}{\kappa^2 \left(h_0 - 1 \right)\Phi^2}\
& \mbox{when}\ h_0 \neq 1 \\
\frac{3}{\kappa^2 \Phi^2}\left( 1 - 2\ln \frac{\Phi}{t_0} \right)\
& \mbox{when}\ h_0 = 1
\end{array} \right. \ ,\\
\label{GBR22}
\xi(\Phi) &=& \left\{ \begin{array}{ll}
\xi_0 - \frac{t_0^{h_0} U_0 \Phi^{3 - h_0} }{8h_0^2 \left( 3 - h_0 \right)}
+ \frac{\Phi^2}{8\kappa^2 \left(h_0 - 1\right)}\ & \mbox{when}\ h_0 \neq 1,\, 3 \\
 - \frac{t_0^3 U_0}{72} \ln \frac{\Phi}{t_2}
+ \frac{\Phi^2}{16\kappa^2}\ & \mbox{when}\ h_0 = 3 \\
\xi_0 - \left(\frac{1}{2}\ln \frac{\Phi}{t_1} + \frac{1}{4}\right)\Phi^2
\ & \mbox{when}\ h_0 = 1
\end{array} \right. \ .
\eea
Here, $\xi_0$ and $t_2$ are integration constants but
they are irrelevant to the action because the Gauss-Bonnet invariant is a
total derivative and $\xi_0$ and $t_2$ correspond to the constant shift of
the coefficient of the Gauss-Bonnet invariant.
We further redefine the scalar field $\phi$ as $\Phi\equiv t_s - \phi$.
The obtained forms of $V(\Phi)$ and $\xi(\Phi)$ do not contain the
parameter $t_s$ corresponding to the Big Rip time. Hence, the resulting
$F({\cal G})$ does not contain $t_s$ either and $t_s$ could be determined
dynamically by initial conditions.

When $\beta\neq 0$, $g(\phi)$ is given by
\be
\label{GBR22B}
g(\phi) = \frac{h_0}{\beta -1}\left(t_s - \phi\right)^{1-\beta} + g_0\ ,
\ee
where $g_0$ is a constant of the integration but the constant is irrelevant
and does not appear in the final expressions of $V(\phi)$ and $\xi(\phi)$.
We therefore choose $g_0=0$.
The $U(\phi)$ has the following form:
\be
\label{GBR23}
U(\phi) = \frac{2h_0\beta }{\kappa^2} \int^\phi d\phi_1 \left(t_s - \phi_1\right)^{-1-\beta}
\e^{-\frac{h_0}{\beta - 1}\left(t_s - \phi_1\right)^{1-\beta}}
= \frac{2h_0 \beta}{\kappa^2 \left(\beta - 1\right)} \int^x dx x^{1/(\beta -1)}
\e^{- \frac{h_0}{\beta - 1} x}\ ,
\ee
where
\be
\label{GBR24}
x \equiv \left(t_s - \phi\right)^{1-\beta} \ .
\ee

We now consider the case $\beta>1$, which corresponds to the Type I
singularity. In this case, $x\to \infty$ when $\phi\to t_s$.
When $x$ is large, the following expression can be used:
\be
\label{GBR25}
\int dx x^\alpha \e^{-ax} = - \e^{-ax}\left( \frac{x^\alpha}{a} + \frac{\alpha}{a^2} x^{\alpha - 1}
+ \frac{\alpha(\alpha -1)}{a^3}x^{\alpha -2} + \cdots \right)\ ,\quad
\left(\alpha=\frac{1}{\beta}\ ,\quad a=\frac{h_0}{\beta - 1} \right)\ .
\ee
Keeping the leading order, one finds
\be
\label{GBR26}
U(\phi) \sim - \frac{2\beta}{\kappa^2\left(t_s - \phi\right)}
\e^{-\frac{h_0}{\beta - 1}\left(t_s - \phi\right)^{1-\beta}} \ ,
\ee
and therefore
\be
\label{GBR27}
V(\Phi) \sim \frac{3h_0^2}{\kappa^2} \Phi^{-2\beta}
+ \frac{6\beta}{\kappa^2}\Phi^{-\beta -1}
\sim \frac{3h_0^2}{\kappa^2} \Phi^{-2\beta}\ ,\quad
\xi(\Phi) \sim \frac{\Phi^{2\beta}}{8\kappa^2 h_0^2} + \xi_0\ .
\ee
Here, $\Phi=t_s - \phi$ again and $\xi_0$ is an irrelevant constant of the
integration.

In case of $\beta<1$, which corresponds to the Type II, III, IV
singularities, we find that $x\to 0$ when $\phi\to t_s$. Using the expression
\be
\label{GBR28}
\int dx x^\alpha \e^{-ax} = \frac{1}{\alpha + 1}x^{\alpha + 1} - \frac{a}{\alpha+2}x^{\alpha+2}
+ \frac{a^2}{2!\left(\alpha+3\right)}x^{\alpha+3} - \cdots\ ,
\ee
and only keeping the leading order, we find
\be
\label{GBR29}
U(\phi) = \frac{2h_0}{\kappa^2}\left(t_s - \phi\right)^{-\beta} + U_0\ ,
\ee
where $U_0$ is a constant of the integration.
Hence,
\be
\label{GBR30}
V(\Phi) \sim - \frac{3h_0^2}{\kappa^2} \Phi^{-2\beta} - 3h_0 U_0 \Phi^{-\beta}\ ,\quad
\xi(\Phi) \sim \left\{ \begin{array}{ll}
 - \frac{\Phi^{\beta + 1}}{4\kappa^2 h_0 (\beta + 1)}
 - \frac{U_0 \Phi^{2\beta + 1}}{8h_0^2 (2\beta + 1)} + \xi_0\
& \mbox{when}\ \beta\neq -1,\,-\frac{1}{2} \\
 - \frac{1}{4\kappa^2 h_0}\ln \frac{\Phi}{t_2} + \frac{U_0 \Phi^{- 1}}{8h_0^2} \
& \mbox{when}\ \beta= -1 \\
 - \frac{\Phi^{1/2}}{2\kappa^2 h_0}
 - \frac{U_0}{8h_0^2 }\ln \frac{\Phi}{t_2}\
& \mbox{when}\ \beta = -\frac{1}{2}
\end{array} \right. \ .
\ee
Here, $\xi_0$ and $t_2$ are (irrelevant) constants of the integration.

In case of the scalar Gauss-Bonnet gravity, there is a freedom or ambiguity
in the choice
of $f(\phi)$. For simplicity, we now choose
\be
\label{GBR31}
t=f(\phi) = \kappa^2 \phi\ .
\ee
The equations (\ref{GBR6}) are rewritten as
\bea
\label{GBR32}
V(\varphi) &=& \frac{3}{\kappa^2}g'\left(\varphi\right)^2 - \frac{1}{2\kappa^4}
 - 3g'\left(\varphi)\right) \e^{g\left(\varphi\right)}  U(\varphi) \ ,\nn
\xi(\varphi) &=& \frac{1}{8}\int^\varphi d\varphi_1
\frac{\e^{g\left(\varphi\right)} }{g'(\varphi_1)^2} U(\varphi_1)\ ,\nn
U(\varphi) &\equiv& \int^\varphi d\varphi_1 \e^{-g\left(\varphi_1\right)} \left(\frac{2}{\kappa^2}
g''\left(\varphi_1\right) + \frac{1}{\kappa^4} \right)\ ,
\eea
where $\varphi\equiv \kappa^2\phi$.
By the calculation similar to the case of $F({\cal G})$ gravity, when
$\beta=1$ (the Type I singularity, usual Big Rip), we find
\be
\label{GBR33}
U(\phi) = \left\{
\begin{array}{ll}
U_0 - \frac{2h_0}{\kappa^2{t_0}^{h_0}\left(h_0 -1\right)} \left(t_s - \varphi\right)^{h_0-1}
 - \frac{\left(t_s - \varphi \right)^{h_0 + 1}}{\kappa^4 t_0^{h_0} \left(h_0 + 1\right)}
\ & \mbox{when}\ h_0 \neq 1 \\
\frac{2}{\kappa^2 t_0} \ln \frac{t_s - \varphi}{t_1}
 - \frac{\left(t_s - \varphi \right)^{h_0 + 1}}{\kappa^4 t_0^{h_0} \left(h_0 + 1\right)}
\ & \mbox{when}\ h_0 = 1
\end{array}\right. \ .
\ee
Here, $U_0$ and $t_1$ are constants of the integration again.
Subsequently,
\bea
\label{GBR34}
V(\Phi) &=& \left\{ \begin{array}{ll}
 - \frac{3h_0 t_0^{h_0} U_0}{\Phi^{h_0 + 1}}
+ \frac{3h_0^2 \left(h_0 + 1\right)}{\kappa^2 \left(h_0 - 1 \right)\Phi^2}
+ \frac{2h_0 -1}{2\kappa^4\left(h_0 + 1\right)}
\ & \mbox{when}\ h_0 \neq 1 \\
\frac{3}{\kappa^2 \Phi^2}\left( 1 - 2\ln \frac{\Phi}{t_0} \right)
+ \frac{2h_0 -1}{2\kappa^4\left(h_0 + 1\right)}
\ & \mbox{when}\ h_0 = 1
\end{array} \right. \ ,\\
\label{GBR35}
\xi(\Phi) &=& \left\{ \begin{array}{ll}
\xi_0 - \frac{t_0^{h_0} U_0 \Phi^{3 - h_0} }{8h_0^2 \left( 3 - h_0 \right)}
+ \frac{\Phi^2}{8\kappa^2 \left(h_0 - 1\right)}
+ \frac{\Phi^4}{32h_0^2\left( h_0 + 1 \right)\kappa^4}
\ & \mbox{when}\ h_0 \neq 1,\, 3 \\
 - \frac{t_0^3 U_0}{72} \ln \frac{\Phi}{t_2}
+ \frac{\Phi^2}{16\kappa^2}
+ \frac{\Phi^4}{32h_0^2\left( h_0 + 1 \right)\kappa^4}
\ & \mbox{when}\ h_0 = 3 \\
\xi_0 - \left(\frac{1}{2}\ln \frac{\Phi}{t_1} + \frac{1}{4}\right)\Phi^2
+ \frac{\Phi^4}{32h_0^2\left( h_0 + 1 \right)\kappa^4}
\ & \mbox{when}\ h_0 = 1
\end{array} \right. \ .
\eea
Here, $\Phi=t_s - \varphi$. In terms of $\Phi$, the kinetic term of $\phi$ is
rewritten as
\be
\label{GBR36}
- \frac{1}{2}\partial_\mu \phi \partial^\mu \phi
= - \frac{1}{2\kappa^4}\partial_\mu \Phi \partial^\mu \Phi\ .
\ee
In case of $\beta>1$, one obtains
\bea
\label{GBR37}
&& U(\varphi) \sim - \frac{2\beta}{\kappa^2\left(t_s - \varphi\right)}
\e^{-\frac{h_0}{\beta - 1}\left(t_s - \varphi\right)^{1-\beta}}
+ \frac{\left(t_s - \varphi\right)^\beta}{h_0 \kappa^4}
\e^{-\frac{h_0}{\beta - 1}\left(t_s - \varphi\right)^{1-\beta}} \ , \nn
&& V(\Phi) \sim \frac{3h_0^2}{\kappa^2} \Phi^{-2\beta} - \frac{3}{\kappa^4}
\ ,\quad
\xi(\Phi) \sim \frac{\Phi^{2\beta}}{8\kappa^2 h_0^2} + \xi_0
 - \frac{\Phi^{3\beta + 1}}{8h_0^3\left(3\beta + 1\right)\kappa^4}\ .
\eea
In case of $\beta<1$, we get
\bea
\label{GBR38}
U(\varphi) &=& \frac{2h_0}{\kappa^2}\left(t_s - \varphi\right)^{-\beta} + U_0
 - \frac{t_s - \varphi}{\kappa^4}\ , \nn
V(\Phi) &\sim& - \frac{3h_0^2}{\kappa^2} \Phi^{-2\beta} - 3h_0 U_0 \Phi^{-\beta}
+ \frac{3h_0}{\kappa^4}\Phi^{1-\beta}\ ,\nn
\xi(\Phi) &\sim& \left\{ \begin{array}{ll}
 - \frac{\Phi^{\beta + 1}}{4\kappa^2 h_0 (\beta + 1)}
 - \frac{U_0 \Phi^{2\beta + 1}}{8h_0^2 (2\beta + 1)} + \xi_0
+ \frac{\Phi^{2\beta + 2}}{16h_0^2 \kappa^4 \left(\beta + 1\right)}\
& \mbox{when}\ \beta\neq -1,\,-\frac{1}{2} \\
 - \frac{1}{4\kappa^2 h_0}\ln \frac{\Phi}{t_2} + \frac{U_0 \Phi^{- 1}}{8h_0^2}
+ \frac{\Phi^{2\beta + 2}}{16h_0^2 \kappa^4 \left(\beta + 1\right)}\
& \mbox{when}\ \beta= -1 \\
 - \frac{\Phi^{1/2}}{2\kappa^2 h_0}
 - \frac{U_0}{8h_0^2 }\ln \frac{\Phi}{t_2}
+ \frac{\Phi^{2\beta + 2}}{16h_0^2 \kappa^4 \left(\beta + 1\right)}\
& \mbox{when}\ \beta = -\frac{1}{2}
\end{array} \right. \ .
\eea

Thus, any type of the finite-time future singularities can be realized in
$F({\cal G})$ gravity and the scalar-Gauss-Bonnet one. 
We have constructed the specific examples of the above models containing the
finite-time future singularity by using the reconstruction program.
We should note, however, that the relation between the appearance of the
singularity and the forms of $V(\phi)$ and $\xi(\phi)$ in
the scalar-Gauss-Bonnet gravity is not so clear.
In other words, even if the explicit forms of $V(\phi)$ and $\xi(\phi)$
are given, it is difficult to understand if the theory could generate the
singularity or not.
The corresponding investigation of the asymptotic behavior of solution should
be done in order to answer to this question.
In case of $F({\cal G})$ gravity, however, if $F({\cal G})$ contains the term
like $\sqrt{{\cal G}_0^2 - {\cal G}^2}$
$({\cal G}_0>0)$, which becomes imaginary, and therefore inconsistent, if
$\left|{\cal G}\right|> {\cal G}_0$, the singularity should not appear.
In other words, even if the solution contains the finite-time future
singularity, the additional modification of the action may resolve it.

\section{Non-minimal Maxwell-Einstein gravity \label{Sec6}}

In this section, we consider some cosmological effects in the non-minimal
Maxwell-Einstein gravity with general gravitational coupling.
We account for our model and derive the effective energy density and pressure
of the universe.

\subsection{Model}

We consider the following model action~\cite{Bamba:2008ja}:
\begin{eqnarray}
S_{\mathrm{GR}} \Eqn{=}
\int d^{4}x \sqrt{-g}
\left[ \hspace{1mm}
{\mathcal{L}}_{\mathrm{EH}}
+{\mathcal{L}}_{\mathrm{EM}}
\hspace{1mm} \right]\,,
\label{eq:6.1} \\
{\mathcal{L}}_{\mathrm{EH}}
\Eqn{=} \frac{1}{2\kappa^2} R\,,
\label{eq:6.2} \\
{\mathcal{L}}_{\mathrm{EM}}
\Eqn{=}
 -\frac{1}{4} I(R)
F_{\mu\nu}F^{\mu\nu}\,,
\label{eq:6.3} \\
I(R) \Eqn{=} 1+ \tilde{I}(R)\,,
\label{eq:6.4}
\end{eqnarray}
where $F_{\mu\nu} = {\partial}_{\mu}A_{\nu} - {\partial}_{\nu}A_{\mu}$
is the electromagnetic field-strength tensor. Here, $A_{\mu}$ is the $U(1)$
gauge field. Furthermore, $\tilde{I}(R)$ is an arbitrary function of $R$.
It is known that the coupling between the scalar curvature and
the Lagrangian of the electromagnetic field arises in curved
space-time due to one-loop vacuum-polarization effects in
Quantum Electrodynamics~\cite{Drummond:1979pp}.
(In Ref.~\cite{Bamba:2008xa}, a non-minimal gravitational Yang-Mills (YM)
theory, in which the YM field couples to a function of the scalar curvature,
has been discussed. Note that this study maybe generalized also for
$F(R)$ gravity coupled to non-linear
electrodynamics~\cite{Hollenstein:2008hp}.)

Taking variations of the action Eq.\ (\ref{eq:6.1}) with respect to the metric
$g_{\mu\nu}$ and the $U(1)$ gauge field $A_{\mu}$, we obtain the gravitational
field equation and the equation of motion of $A_{\mu}$ as~\cite{Bamba:2008ja}
\begin{eqnarray}
R_{\mu \nu} - \frac{1}{2}g_{\mu \nu}R
= \kappa^2 T^{(\mathrm{EM})}_{\mu \nu}\,,
\label{eq:6.5}
\end{eqnarray}
with
\begin{eqnarray}
\hspace{-5mm}
T^{(\mathrm{EM})}_{\mu \nu}
\Eqn{=}
I(R) \left( g^{\alpha\beta} F_{\mu\beta} F_{\nu\alpha}
 -\frac{1}{4} g_{\mu\nu} F_{\alpha\beta}F^{\alpha\beta} \right)
\nonumber \\
&&
{}+\frac{1}{2} \biggl\{ I^{\prime}(R)
F_{\alpha\beta}F^{\alpha\beta} R_{\mu \nu}
+ g_{\mu \nu} \Box \left[ I^{\prime}(R)
F_{\alpha\beta}F^{\alpha\beta} \right]
 - {\nabla}_{\mu} {\nabla}_{\nu}
\left[ I^{\prime}(R)
F_{\alpha\beta}F^{\alpha\beta} \right]
\biggr\}
\,,
\label{eq:6.6}
\end{eqnarray}
and
\begin{eqnarray}
 -\frac{1}{\sqrt{-g}}{\partial}_{\mu}
\left( \sqrt{-g} I(R) F^{\mu\nu}
\right) = 0\,,
\label{eq:6.7}
\end{eqnarray}
respectively,
where
$T^{(\mathrm{EM})}_{\mu \nu}$ is the contribution to
the energy-momentum tensor from the electromagnetic field.

\subsection{Effective energy density and pressure of the universe}

We now assume the flat FRW space-time with the metric in Eq.~(\ref{eq:KB1}).
We here consider the case in which there exist only magnetic fields and
hence electric fields are negligible.
In addition, only one component of $\Vec{B}$ is non-zero
and hence other two components are zero.
In this case, it follows from $\mathrm{div} \Vec{B} = 0$ that
the off-diagonal components of the last term
on the r.h.s. of Eq.\ (\ref{eq:6.6}) for
$T^{(\mathrm{EM})}_{\mu \nu}$, i.e.,
${\nabla}_{\mu} {\nabla}_{\nu} \left[ f^{\prime}(R)
F_{\alpha\beta}F^{\alpha\beta} \right]$
are zero.
Thus, all of the off-diagonal components of $T^{(\mathrm{EM})}_{\mu \nu}$
are zero (for the argument about the problem of off-diagonal components of
electromagnetic energy-momentum tensor in non-minimal Maxwell-gravity
theory, see~\cite{OD}).
Moreover, because we assume that there exist the magnetic fields as background
quantities at the 0th order, the magnetic fields do not have the dependence on
the space components $\Vec{x}$.

In the FRW background, the equation of motion for the
$U(1)$ gauge field in the Coulomb gauge, ${\partial}^jA_j(t,\Vec{x}) =0$,
and the case of $A_{0}(t,\Vec{x}) = 0$, becomes
\begin{eqnarray}
{\ddot{A}}_i(t,\Vec{x})
+ \left( H + \frac{\dot{I}}{I}
\right) {\dot{A}}_(t,\Vec{x})
 - \frac{1}{a^2}\Lap\, A_i(t,\Vec{x}) = 0\,,
\label{eq:6.8}
\end{eqnarray}
where $\Lap = {\partial}^i {\partial}_i$ is the flat 3-dimensional Laplacian.
It follows from Eq.~(\ref{eq:6.8}) that the Fourier mode $A_i(k,t)$
satisfies the equation
\begin{eqnarray}
{\ddot{A}}_i(k,t) + \left( H + \frac{\dot{I}}{I} \right)
               {\dot{A}}_i(k,t) + \frac{k^2}{a^2} A_i(k,t) = 0\,.
\label{eq:6.9}
\end{eqnarray}
Replacing the independent variable $t$ by conformal time
$\eta = \int dt/a(t)$, we find that Eq.~(\ref{eq:6.9}) becomes
\begin{eqnarray}
\frac{\partial^2 A_i(k,\eta)}{\partial \eta^2} +
\frac{1}{I(\eta)} \frac{d I(\eta)}{d \eta}
\frac{\partial A_i(k,\eta)}{\partial \eta}
+ k^2 A_i(k,\eta) = 0\,.
\label{eq:6.10}
\end{eqnarray}

It is impossible to obtain the exact solution of Eq.~(\ref{eq:6.10})
for the generic evolution of the coupling function $I$ at the inflationary
stage. However, by using the WKB approximation on subhorizon scales and the
long-wavelength approximation on superhorizon scales, and matching these
solutions at the horizon crossing~\cite{Bamba-mag-2}, we find an
approximate solution as
\begin{eqnarray}
\hspace{-5mm}
\left|A_i(k,\eta)\right|^2
= |\bar{C}(k)|^2
= \frac{1}{2kI(\eta_k)}
\left|1- \left[ \frac{1}{2}\frac{1}{kI(\eta_k)}\frac{d I(\eta_k)}{d \eta}
+ i \right]k\int_{\eta_k}^{{\eta}_{\mathrm{f}}}
\frac{I(\eta_k)}{I \left(\Tilde{\Tilde{\eta}} \right)}
d\Tilde{\Tilde{\eta}}\,\right|^2\,,
\label{eq:6.11}
\end{eqnarray}
where $\eta_k$ and ${\eta}_{\mathrm{f}}$ are the conformal time
at the horizon-crossing and one at the end of inflation,
respectively. From Eq.~(\ref{eq:6.11}), we obtain the amplitude of
the proper magnetic fields in the position space
\begin{eqnarray}
|{B_i}^{(\mathrm{proper})}(t)|^2 =
\frac{k|\bar{C}(k)|^2}{\pi^2}\frac{k^4}{a^4}\,,
\label{eq:6.12}
\end{eqnarray}
on a comoving scale $L=2\pi/k$.
Thus, from Eq.~(\ref{eq:6.12}) we see that the proper magnetic fields evolves
as $|{B_i}^{(\mathrm{proper})}(t)|^2 = |\bar{B}|^2/a^4$, where
$|\bar{B}|$ is a constant.
(The validity of this behavior of the proper magnetic fields, namely,
that $|\bar{B}|$ is a constant, is shown in Appendix.)
This means that the influence of the coupling function $I$ on
the value of the proper magnetic fields exists only during inflation.
(On the other hand, because the expression of the energy density of
the magnetic fields is described by that of the magnetic fields multiplying
$I$ due to the Lagrangian (\ref{eq:6.3}), the energy density of the magnetic
fields depends on $I$ also after inflation. We can see this point from
the first term on the r.h.s. of Eq.~(\ref{eq:6.13}) shown below.)
The conductivity of the universe ${\sigma}_\mathrm{c}$
is negligibly small during inflation, because there are few charged particles
at that time. After the reheating stage, a number of charged particles are
produced, so that the conductivity immediately jumps to a large
value:~${\sigma}_\mathrm{c} \gg H$.
Consequently, for a large enough conductivity at the reheating stage,
the proper magnetic fields evolve in proportion to $a^{-2}(t)$ in the
radiation-dominated stage and the subsequent matter-dominated
stage~\cite{Ratra:1991bn}.

In this case, it follows from Eq.~(\ref{eq:6.6}) that
the quantity corresponding to the effective energy density of the universe
$\rho_\mathrm{eff}$ and that corresponding to the effective pressure
$p_\mathrm{eff}$ are given by
\begin{eqnarray}
\hspace{-15mm}
&&
\rho_\mathrm{eff} =
\left\{ \frac{I(R)}{2} + 3\left[
 -\left( 5 H^2 + \dot{H} \right) I^{\prime}(R) +
6 H \left( 4H\dot{H} + \ddot{H} \right) I^{\prime\prime}(R)
\right]
\right\}
\frac{|\bar{B}|^2}{a^4}\,,
\label{eq:6.13} \\
\hspace{-15mm}
&&
p_\mathrm{eff} = \biggl[ -\frac{I(R)}{6} +
\left( -H^2 + 5\dot{H} \right) I^{\prime}(R) -
6 \left( -20H^2\dot{H} + 4\dot{H}^2 - H\ddot{H} + \dddot{H} \right)
I^{\prime\prime}(R) \nonumber \\
&& \hspace{60mm}
{}-36\left( 4H\dot{H} + \ddot{H} \right)^2 I^{\prime\prime\prime}(R)
\biggr]
\frac{|\bar{B}|^2}{a^4}\,,
\label{eq:6.14}
\end{eqnarray}
where we have used the following relations by taking into account
that electric fields are negligible:
$
g^{\alpha\beta} F_{0\beta} F_{0\alpha}
 -\left(1/4\right) g_{00} F_{\alpha\beta}F^{\alpha\beta}
= |{B_i}^{(\mathrm{proper})}(t)|^2/2,
$
and
$
F_{\alpha\beta}F^{\alpha\beta}
= 2|{B_i}^{(\mathrm{proper})}(t)|^2
$.

Finally, we remark the following point.
Suppose that $I(R)$ is (almost) constant at the present time.
We now assume that for the small curvature, $I(R)$ behaves as
\be
I(R) \sim I_0 R^\alpha \ ,
\label{eq:7.5}
\ee
with constants $I_0$ and $\alpha$. Here, we consider the case $\alpha < 0$.
The energy density of the magnetic fields is given by
$\rho_B = \left(1/2\right) |{B_i}^{(\mathrm{proper})}(t)|^2 I(R) =
\left[|\bar{B}|^2/\left(2a^4\right)\right]I(R)$.
Here we take de Sitter background as the future universe.
In such a case, when $R$ tends to zero in the future, the energy density of
the magnetic field becomes larger and larger in comparison with
its current value.
Hence, if for the small curvature, non-minimal gravitational coupling of
the electromagnetic fields behaves as $\sim I_0 R^\alpha$ with
$\alpha < 0$ in the future, the strength of current magnetic fields of
the universe may evolve to very large values in the future universe.

\section{Finite-time future singularities in non-minimal Maxwell-Einstein
gravity \label{Sec7}}

In this section, we investigate the finite-time future singularities
in non-minimal Maxwell-Einstein gravity.
We study the forms of the non-minimal gravitational coupling of the
electromagnetic field generating the finite-time future singularities
and the general conditions for the non-minimal gravitational coupling of
the electromagnetic field in order that the finite-time future singularities
cannot emerge.

It follows from (\ref{GBR18}) that the FRW equations are given by
\begin{eqnarray}
\frac{3}{\kappa^2}H^2\ \Eqn{=} \rho_\mathrm{eff}\,,
\label{eq:7.1} \\
-\frac{1}{\kappa^2}\left(2\dot H + 3H^2\right) \Eqn{=} p_\mathrm{eff}\,,
\label{eq:7.2}
\end{eqnarray}
where $\rho_\mathrm{eff}$ and $p_\mathrm{eff}$ are given by
Eqs.~(\ref{eq:6.13}) and (\ref{eq:6.14}), respectively.

We investigate the form of $I(R)$ which produces the Big Rip singularity,
\be
H \sim \frac{h_0}{t_0 - t}\ ,
\label{eq:7.3}
\ee
where $h_0$ is a positive constant, and $H$ diverges at $t = t_0$.
In this case,
\be
R \sim \frac{12h_0^2 + 6h_0}{\left(t_0 - t\right)^2}\ ,
\quad a \sim a_0 \left(t_0 - t\right)^{- h_0}\ ,
\label{eq:7.4}
\ee
where $a_0$ is a constant.
We now assume that for the large curvature, $I(R)$ behaves as
Eq.~(\ref{eq:7.5}).
Hence $\rho_{\rm eff}$ in Eq.~(\ref{eq:6.13}) behaves as
$\left(t_0 - t\right)^{-2\alpha + 4h_0}$, but the l.h.s. in
the FRW equation $3H^2/\kappa^2 = \rho_{\rm eff}$
does as $\left(t_0 - t\right)^{-2}$.
The consistency gives
\be
 - 2 = -2 \alpha + 4h_0\ ,
\label{eq:7.6}
\ee
namely
\be
h_0 = \frac{\alpha -1 }{2} \quad \mbox{or}\quad \alpha= 1 + 2h_0\ .
\label{eq:7.7}
\ee
Eq.~(\ref{eq:7.1}) also shows
\bea
\frac{3h_0^2}{\kappa^2} \Eqn{=}  I_0 \left( 12h_0^2 + 6h_0 \right)^{\alpha -2}
\left\{ \frac{\left( 12h_0^2 + 6h_0 \right)^2}{2}
+ 3\left[ - \alpha \left( 12h_0^2 + 6h_0 \right)\left( h_0 + 5h_0^2 \right)
+ 6\alpha \left( \alpha -1 \right) h_0 \left( 2h_0 + 4 h_0^2 \right)\right]
\right\} \frac{|\bar{B}|^2}{a_0^4} \nn
\Eqn{=} - \frac{I_0 h_0 \left( 12h_0^2 + 6h_0 \right)^\alpha
|\bar{B}|^2}{2 a_0^4} \ ,
\label{eq:7.8}
\eea
which requires that $I_0$ should be negative. In the second line of
(\ref{eq:7.8}), we have used (\ref{eq:7.7}) and deleted $\alpha$.

As a result,
it follows from Eqs.~(\ref{eq:7.5}) and (\ref{eq:7.7}) that
the Big Rip singularity in Eq.~(\ref{eq:7.3}) can emerge only when for the
large curvature, $I(R)$ behaves as $R^{1+2h_0}$. If the form of $I(R)$ is
given by other terms, the Big Rip singularity cannot emerge.
We here note that if exactly $I(R) = I_0 R^\alpha$,
$H=h_0/\left(t_0 - t\right)$ is an exact solution.

Next, we study the form of $I(R)$ which gives a more general singularity in
Eq.~(\ref{frlv9}).
In this case,
\begin{eqnarray}
R \sim 6h_0 \left[\beta + 2h_0 \left( t_0 -t \right)^{-\left(\beta-1 \right)}
\right] \left( t_0 -t \right)^{-\left(\beta+1 \right)}\,,
\quad a \sim a_0 \exp \left[ \frac{h_0}{\beta-1}
\left( t_0 -t \right)^{-\left(\beta -1 \right)} \right]\,.
\label{eq:7.9}
\end{eqnarray}

We also assume that
for the large curvature, $I(R)$ behaves as Eq.~(\ref{eq:7.5}).
If $\beta <-1$, in the limit $t \to t_0$, $R \to 0$. Hence,
we consider this case later.
If $\beta >1$, $a \to \infty$, and hence
$\rho_\mathrm{eff} \to 0$ and $p_\mathrm{eff} \to 0$
because $\rho_\mathrm{eff} \propto a^{-4}$ and
$p_\mathrm{eff} \propto a^{-4}$.
On the other hand, $H \to \infty$.
Thus Eqs.~(\ref{eq:7.1}) and (\ref{eq:7.2}) cannot be satisfied.

If $\alpha > 0$ and $0< \beta <1$, $\rho_{\rm eff}$ in Eq.~(\ref{eq:6.13})
evolves as $\left(t_0 - t\right)^{-\alpha \left( \beta + 1 \right)}$, but the
l.h.s. of Eq.~(\ref{eq:7.1}) does as $\left(t_0 - t\right)^{-2\beta}$.
Thus, the consistency gives
\begin{eqnarray}
- 2\beta = -\alpha \left( \beta + 1 \right)\,,
\label{eq:7.10}
\end{eqnarray}
namely,
\begin{eqnarray}
\beta = \frac{\alpha}{2-\alpha} \quad \mbox{or} \quad
\alpha = \frac{2\beta}{\beta+1}\,.
\label{eq:7.11}
\end{eqnarray}
 From Eq.~(\ref{eq:7.1}), we also find
\begin{eqnarray}
\frac{3h_0^2}{\kappa^2} =
- \frac{I_0 \left( 6h_0 \beta \right)^\alpha \left( 1-\beta \right)
|\bar{B}|^2}{2 a_0^4 \left( \beta + 1 \right)}\,,
\label{eq:7.12}
\end{eqnarray}
where we have used Eq.~(\ref{eq:7.11}), and on the l.h.s. we have taken only
the leading term.
Eq.~(\ref{eq:7.12}) requires that $I_0$ should be negative.
Consequently, if $\alpha > 0$ and $0< \beta <1$,
in the limit $t \to t_0$,
$a \to a_0$, $R \to \infty$,
$\rho_\mathrm{eff} \to \infty$, and
$|p_\mathrm{eff}| \to \infty$. Hence the Type III singularity emerges.
If $\alpha > 0$ and $-1< \beta <0$, $\rho_\mathrm{eff} \to \infty$,
but $H \to 0$. Hence Eq.~(\ref{eq:7.1}) cannot be satisfied.

If $\left( \beta-1 \right)/\left( \beta+1 \right) < \alpha <0$ and
$-1< \beta <0$, in the limit $t \to t_0$,
$a \to a_0$, $R \to \infty$, $\rho_\mathrm{eff} \to 0$, and
$|p_\mathrm{eff}| \to \infty$.
Although the final value of $\rho_\mathrm{eff}$ is not a finite one but
vanishes, this singularity can be considered to the Type II.
The reason is as follows.
In this case, when $I$ and $H$ are given by
$I = 1 + I_0 R^\alpha$ and $H=H_0+h_0 \left(t_0 -t\right)^{-\beta}$,
where $H_0$ is a constant,
respectively, in the above limit $\rho_\mathrm{eff} \to \rho_0$. From
Eqs.~(\ref{eq:6.13}) and (\ref{eq:7.1}), we find
$\rho_0 = 3H_0^2/\kappa^2 =
|\bar{B}|^2/\left(2 a_0^4\right)$. Hence, $\rho_0$ is a finite value.

If $\alpha \leq \left( \beta-1 \right)/\left( \beta+1 \right)$ and
$-1< \beta <0$, in the limit $t \to t_0$,
$a \to a_0$, $R \to \infty$,
$\rho_\mathrm{eff} \to 0$, and $|p_\mathrm{eff}| \to 0$,
but $\dot{H} \to \infty$.
Hence Eq.~(\ref{eq:7.2}) cannot be satisfied.
If $\alpha < 0$ and $0< \beta <1$, $\rho_\mathrm{eff} \to 0$, but
$H \to \infty$. Hence Eq.~(\ref{eq:7.1}) cannot be satisfied.

In addition, we investigate the case in which $\beta <-1$. In this case,
in the limit $t \to t_0$, $a \to a_0$ and $R \to 0$.
We assume that for the small curvature, $R$ behaves as
Eq.~(\ref{eq:7.5}).
If $\alpha \geq \left( \beta-1 \right)/\left( \beta+1 \right)$,
in the limit $t \to t_0$, $\rho_\mathrm{eff} \to 0$,
$|p_\mathrm{eff}| \to 0$, and higher derivatives of $H$
diverge. Hence the Type IV singularity emerges.
If $0 < \alpha < \left( \beta-1 \right)/\left( \beta+1 \right)$,
$\rho_\mathrm{eff} \to 0$ and $|p_\mathrm{eff}| \to \infty$.
However, $H \to 0$ and $\dot{H} \to 0$. Hence Eq.~(\ref{eq:7.2}) cannot be
satisfied.

We note that
if $I(R)$ is a constant (the case in which $I(R)=1$ corresponds to the
ordinary Maxwell theory), any singularity cannot emerge.

Here we mention the case in which $I(R)$ is given by the Hu-Sawicki
form~\cite{HS}
\begin{eqnarray}
I(R) = I_{\mathrm{HS}}(R) \equiv \frac{c_1 \left(R/m^2 \right)^n}
{c_2 \left(R/m^2 \right)^n + 1}\,,
\label{eq:7.13}
\end{eqnarray}
which satisfies the conditions:
$
\lim_{R\to\infty} I_{\mathrm{HS}}(R) = c_1/c_2 = \mbox{const}
$
and
$
\lim_{R\to 0} I_{\mathrm{HS}}(R) = 0.
$
Here, $c_1$ and $c_2$ are dimensionless constants, $n$ is a positive
constant, and $m$ denotes a mass scale.
The following form~\cite{Nojiri:2007as} also has the same features:
\begin{eqnarray}
I(R) = I_{\mathrm{NO}}(R) \equiv
\frac{\left[ \left(R/M^2\right) - \left(R_\mathrm{c}/M^2\right) \right]^{2q+1}
+ {\left(R_\mathrm{c}/M^2\right)}^{2q+1}}
{c_3 + c_4 \left\{
\left[ \left(R/M^2\right) - \left(R_\mathrm{c}/M^2\right) \right]^{2q+1} +
{\left(R_\mathrm{c}/M^2\right)}^{2q+1} \right\}}\,,
\label{eq:7.14}
\end{eqnarray}
which satisfies the conditions:
$
\lim_{R\to\infty} I_{\mathrm{NO}}(R) = 1/c_4 = \mbox{const}
$
and
$
\lim_{R\to 0} I_{\mathrm{NO}}(R) = 0.
$
Here, $c_3$ and $c_4$ are dimensionless constants,
$q$ is a positive integer, $M$ denotes a mass scale, and $R_\mathrm{c}$ is
current curvature.
If $\beta < -1$ and
$I(R)$ is given by $I_{\mathrm{HS}}(R)$ in Eq.~(\ref{eq:7.13}) or
$I_{\mathrm{NO}}(R)$ in Eq.~(\ref{eq:7.14}), in the limit $t \to t_0$,
$a \to a_0$, $R \to 0$, $\rho_\mathrm{eff} \to 0$, and
$|p_\mathrm{eff}| \to 0$.
In addition, higher derivatives of $H$ diverge.
Thus the Type IV singularity emerges.

Consequently, it is demonstrated that Maxwell theory which is coupled
non-minimally with Einstein gravity may produce finite-time singularities
in future, depending on the form of non-minimal gravitational coupling.

The general conditions for $I(R)$ in order that the finite-time future
singularities whose form is given by Eqs.~(\ref{eq:7.3}) or (\ref{frlv9})
cannot emerge are that in the limit $t \to t_0$, $I(R) \to \bar{I}$,
where $\bar{I} (\neq 0)$ is a finite constant,
$I^{\prime}(R) \to 0$, $I^{\prime\prime}(R) \to 0$, and
$I^{\prime\prime\prime}(R) \to 0$.

\section{Influence of non-minimal gravitational coupling on the
finite-time future singularities in modified gravity \label{Sec8}}

In this section,
we consider the case in which there exist the finite-time future singularities
in modified $F(R)$ gravity and investigate the influence of non-minimal
gravitational coupling on them.
We show that a non-minimal gravitational coupling of the electromagnetic
field can remove the finite-time future singularities or
make the singularity stronger (or weaker).

In this case,
the total energy density and pressure of the universe are given by
$\rho_{\mathrm{tot}} = \rho_\mathrm{eff} + \rho_{\mathrm{MG}}$
and $p_{\mathrm{tot}} = p_\mathrm{eff} + p_{\mathrm{MG}}$, respectively.
Here,
$\rho_\mathrm{eff}$ and $p_\mathrm{eff}$ are given by
Eqs.~(\ref{eq:6.13}) and (\ref{eq:6.14}), respectively.
Moreover, it follows from Eqs.~(\ref{Cr4}) and (\ref{Cr4b}) that
$\rho_{\mathrm{MG}}$ and $p_{\mathrm{MG}}$ are given by
\begin{eqnarray}
&& \hspace{-15mm}
\rho_{\mathrm{MG}} =
\frac{1}{\kappa^2}\left[-\frac{1}{2}f(R) + 3\left(H^2  + \dot H\right) f'(R)
 - 18 \left(4H^2 \dot H + H \ddot H\right)f''(R)\right]\,,
\label{eq:8.1} \\
&& \hspace{-15mm}
p_{\mathrm{MG}} =
\frac{1}{\kappa^2}\left[\frac{1}{2}f(R) - \left(3H^2 + \dot H \right)f'(R)
+ 6 \left(8H^2 \dot H + 4{\dot H}^2
+ 6 H \ddot H + \dddot H \right)f''(R) + 36\left(4H\dot H + \ddot H\right)^2
f'''(R) \right]\,.
\label{eq:8.2}
\end{eqnarray}

In this case,
it follows from Eqs.~(\ref{GBR18}), (\ref{eq:6.13}), (\ref{eq:6.14}),
(\ref{eq:8.1}), and (\ref{eq:8.2}) that the FRW equations are given by
\begin{eqnarray}
&& \hspace{-10mm}
\frac{3}{\kappa^2}H^2\ = \rho_\mathrm{tot}
= \left\{ \frac{I(R)}{2} + 3\left[
-\left( 5H^2 + \dot{H} \right) I^{\prime}(R) +
6 H \left( 4H\dot{H} + \ddot{H} \right) I^{\prime\prime}(R)
\right] \right\}
\frac{|\bar{B}|^2}{a^4} \nonumber \\
&& \hspace{16mm}
{}+\frac{1}{\kappa^2}\left[-\frac{1}{2}\left( F(R)-R \right) +
3\left(H^2  + \dot H\right) \left( F^{\prime}(R)-1 \right)
 -18 \left(4H^2 \dot H + H \ddot H\right) F^{\prime\prime}(R)\right]\,,
\label{eq:8.3} \\
&& \hspace{-10mm}
 -\frac{1}{\kappa^2}\left(2\dot H + 3H^2\right) = p_\mathrm{tot} =
\biggl[ -\frac{I(R)}{6} +
\left( - H^2 + 5\dot{H} \right) I^{\prime}(R) -
6 \left( -20H^2\dot{H} + 4\dot{H}^2 - H\ddot{H} + \dddot{H}
\right) I^{\prime\prime}(R) \nonumber \\
&& \hspace{35mm}
 -36\left( 4H\dot{H} + \ddot{H} \right)^2 I^{\prime\prime\prime}(R)
\biggr]
\frac{|\bar{B}|^2}{a^4}
+\frac{1}{\kappa^2} \biggl[ \frac{1}{2}\left( F(R)-R \right)
 -\left(3H^2 + \dot H \right) \left( F^{\prime}(R)-1 \right) \nonumber \\
&& \hspace{35mm}
+ 6 \left(8H^2 \dot H + 4{\dot H}^2
+ 6 H \ddot H + \dddot H \right)F^{\prime\prime}(R) +
36\left(4H\dot H + \ddot H\right)^2 F^{\prime\prime\prime}(R) \biggr]\,.
\label{eq:8.4}
\end{eqnarray}

Using Eqs.~(\ref{eq:8.3}) and (\ref{eq:8.4}), we find
\begin{eqnarray}
0 \Eqn{=}
\biggl[ \frac{I(R)}{3} + 2\left(-8H^2 + \dot{H} \right) I^{\prime}(R) +
6\left(32 H^2 \dot{H} -4\dot{H}^2 + 4H\ddot{H} -\dddot{H} \right)
I^{\prime\prime}(R)
 -36\left(4H\dot{H} + \ddot{H} \right)^2 I^{\prime\prime\prime}(R)
\biggr] \frac{|\bar{B}|^2}{a^4} \nonumber \\
&&
{}+\frac{1}{\kappa^2} \left[ 2\dot{H}F^{\prime}(R)
+6\left( -4H^2 \dot{H} +4\dot{H}^2 + 3H\ddot{H} + \dddot{H} \right)
F^{\prime\prime}(R)
+36\left(4H\dot H + \ddot H\right)^2 F^{\prime\prime\prime}(R)
\right]\,.
\label{eq:8.5}
\end{eqnarray}

In modified gravity with the ordinary Maxwell theory,
if $F(R)$ is given by Eq.~(\ref{RR2}), i.e.,
$F(R) \sim F_0 R + F_1 R^q$, where $q$ is a constant,
a finite-time future singularity appears.
Let us account for non-minimal gravitational electromagnetic theory in this
$F(R)$-gravity model.

When $H$ behaves as
\begin{eqnarray}
H \sim h_0 \left( t_0 -t \right)^u\,,
\label{eq:8.6}
\end{eqnarray}
where $u$ is an positive integer, there exists no
finite-time future singularity. In what follows,
we consider the case $u \geq 2$. From Eq.~(\ref{eq:8.6}),
we find
\begin{eqnarray}
R \sim 6\dot{H}
\sim -6u h_0 \left( t_0 -t \right)^{u-1}\,,
\quad
a \sim a_0 \exp \left[-\frac{h_0}{u+1}
\left( t_0 -t \right)^{u+1} \right]\,,
\label{eq:8.7}
\end{eqnarray}
where in the expression of $R$ we have taken only the leading term.

We investigate the form of the non-minimal gravitational coupling of
the electromagnetic field $I(R)$ which produces the solution (\ref{eq:8.7}).
We here assume that $I(R)$ behaves as (\ref{eq:7.5}).
The first and second terms on the r.h.s. of Eq.~(\ref{eq:8.5}) are
the non-minimal gravitational electromagnetic coupling and the
modified-gravity sectors, respectively.
When $t$ is close to $t_0$,
the leading term of the non-minimal gravitational electromagnetic coupling
sector evolves as
$\left( t_0 -t \right)^{\left( u-1 \right) \left( \alpha-1 \right) -2}$.
On the other hand, if $q \leq 1$, or $q>1$ and $u < q/\left(q-2 \right)$,
that of the modified-gravity sector does as
$\left( t_0 -t \right)^{\left( u-1 \right) \left( q-1 \right) -2}$.
Hence the consistency gives $\alpha = q$.
Moreover, in order that the leading term of the non-minimal gravitational
electromagnetic coupling sector should not diverge in the limit $t \to t_0$,
$\alpha$ must be $\alpha \geq \left(u+1\right)/\left( u-1 \right)$.
Taking only the leading terms in Eq.~(\ref{eq:8.5}) and using $\alpha = q$,
we find
\begin{eqnarray}
I_0 = \frac{a_0^4 F_1}{|\bar{B}|^2 \kappa^2}\,.
\label{eq:8.8}
\end{eqnarray}
If $q>1$ and $u \geq q/\left(q-2 \right)$, the leading term of
the modified-gravity sector behaves as
$\left( t_0 -t \right)^{u-1}$.
Hence the consistency gives $\alpha = 2u/\left( u-1 \right)$.
In this case,
taking only the leading terms in Eq.~(\ref{eq:8.5}) and
using $\alpha = 2u/\left( u-1 \right)$, we obtain
\begin{eqnarray}
I_0 = \frac{a_0^4 F_0}{|\bar{B}|^2 \kappa^2}
\frac{u-1}{6 u^2 \left( u+1 \right)}
\left(-6h_0 u \right)^{-2/\left(u-1 \right)}\,.
\label{eq:8.9}
\end{eqnarray}
Consequently, we see that
the non-minimal gravitational coupling of the electromagnetic field
$I(R) \sim I_0 R^\alpha$ with the specific values of $I_0$ and $\alpha$
stated above can resolve the finite-time future singularities which occur
in pure modified gravity.

Next, we study the case in which
non-minimal gravitational coupling of the electromagnetic field
does not remove the singularity but makes it stronger (or weaker).
We also assume that for the large curvature, $I(R)$ behaves as (\ref{eq:7.5}).
Using the result in Eq.~(\ref{rlv13}), we consider the case in which
for the large curvature, $F(R)$ behaves as
$F(R) \propto R^{\bar{q}}$, where $\bar{q} \equiv 1 - \alpha_-/2 <1$.
In this case, the Big Rip singularity in Eq.~(\ref{eq:7.3}) emerges.
It follows from Eq.~(\ref{eq:7.7}) that
if $\alpha =1 + 2h_0$, in the limit $t \to t_0$,
$\rho_\mathrm{eff} \to \infty$ and
$|p_\mathrm{eff}| \to \infty$. Hence, the non-minimal gravitational coupling
of the electromagnetic field makes the singularity stronger.

When there exist a more general singularity in Eq.~(\ref{frlv9}) with
$0<\beta<1$, which is the Type III singularity
and can appear for the form of $F(R)$ in Eq.~(\ref{rlv31C}),
and $\alpha > 0$,
in the limit $t \to t_0$,
$\rho_\mathrm{eff} \to \infty$ and
$|p_\mathrm{eff}| \to \infty$. Thus the non-minimal gravitational coupling
of the electromagnetic field makes the singularity stronger.
If $-1<\beta<0$, namely, there exists the Type II singularity, which can
appear for the form of $F(R)$ in Eq.~(\ref{rlv35}),
and $\left( \beta-1 \right)/\left( \beta+1 \right) < \alpha <0$,
$\rho_\mathrm{eff} \to 0$ and $|p_\mathrm{eff}| \to \infty$.
For $|p_{\mathrm{tot}}| > |p_{\mathrm{MG}}|$
($|p_{\mathrm{tot}}| < |p_{\mathrm{MG}}|$), therefore, the non-minimal
gravitational coupling of the electromagnetic field
makes the singularity stronger (weaker).
Thus, the non-minimal gravitational coupling in Maxwell theory may
qualitatively influence to the universe future. For instances, for some
forms of non-minimal gravitational coupling it resolves the finite-time
future singularity or it may change its properties.

\section{Conclusion \label{Sec9}}

In the present paper, we have considered the finite-time future singularities
in modified gravity: $F(R)$ gravity, scalar-Gauss-Bonnet or modified
Gauss-Bonnet one and effective fluid with the inhomogeneous EoS.
It has been demonstrated that depending on the specific form of the model
under consideration, the universe may approach to the finite-time future
singularity of all four known types.
It is not easy to say from the very beginning if the particular theory brings
the accelerating universe into the singularity or not.
As a rule, each theory should be checked to see if the singularity appears
on the FRW accelerating solutions. It is also interesting that the finite-time
future singularity cannot be seen through the study of local tests:
theory passing known local and cosmological tests may produce
the future universe with (or without) the singularity.
We have presented several examples of modified gravity having the accelerating
dark energy solutions with the finite-time future singularity by explicitly
using the reconstruction program.
Our results are shown in the Table~\ref{tableI} below, where the possibility
of four types of the finite-time future singularity is explained for modified
gravities under consideration.

\begin{center}

\begin{table}[b]
\caption{\label{tableI} Summary of the results: $\rho_0$, $c$, $c_1$, $c_2$, $c_3$, $\tilde c_1$,
$\tilde c_2$, and $\tilde c_3$ are proper constants.}

{%
\footnotesize
\begin{tabular}{|p{3.1cm}|p{3.9cm}|p{3.9cm}|p{3.9cm}|p{3.9cm}|}
\hline
& \multicolumn{1}{c|}{Type I}
& \multicolumn{1}{c|}{Type II}
& \multicolumn{1}{c|}{Type III}
& \multicolumn{1}{c|}{Type IV} \\
& \multicolumn{1}{c|}{$\left(\beta\geq 1\right)$}
& \multicolumn{1}{c|}{$\left(-1<\beta<0\right)$}
& \multicolumn{1}{c|}{$\left(0<\beta<1\right)$}
& \multicolumn{1}{c|}{$\left(\beta<-1,\,\beta:\mbox{not integer}\right)$} \\
\hline
Einstein gravity with inhomogeneous EoS
& \multicolumn{2}{c|}{$p+\rho \propto \rho^{(\beta + 1)/2\beta}$}
& \multicolumn{2}{c|}{$p+\rho \propto \left(\rho^{1/2} - \rho_0^{1/2}\right)^{(\beta + 1)/\beta}$} \\
\hline
$F(R)$ gravity
& \multicolumn{1}{c|}{$F(R)\propto \e^{cR^{\frac{\beta -1}{2\beta}}} R^{-\frac{1}{4}}$}
& \multicolumn{1}{c|}{$F(R)\propto \e^{-cR^{\frac{\beta -1}{\beta+1}}}
R^{-\frac{\beta^2 + 2\beta + 9}{8(\beta +1)}}$}
& \multicolumn{1}{c|}{$F(R)\propto \e^{-cR^{\frac{\beta -1}{\beta+1}}} R^{\frac{7}{4}}$}
& \multicolumn{1}{c|}{$F(R)\propto \e^{-cR^{\frac{\beta -1}{\beta+1}}}
R^{-\frac{\beta^2 + 2\beta + 9}{8(\beta +1)}}$} \\
\hline
Scalar-Gauss-Bonnet gravity
& \multicolumn{1}{c|}{$V(\Phi) \propto \Phi^{-2\beta}$, $\xi(\Phi) \propto \Phi^{2\beta}$}
& \multicolumn{3}{c|}{$V(\Phi)\propto \Phi^{-\beta}$, $\xi(\Phi) \propto \left\{
\begin{array}{ll} c_1 \Phi^{\beta+1} + c_2 \Phi^{2\beta + 2}\quad & \left(\beta\neq -1,-1/2\right) \\
c_1 \ln \Phi + c_2 \Phi^{-1} & \left(\beta=-1\right) \\
c_1 \Phi^{1/2} + c_2 \ln \Phi & \left(\beta=-1/2\right)
\end{array}\right. $} \\
\hline
Modified Gauss-Bonnet gravity
&  \multicolumn{1}{c|}{\parbox[c]{3.4cm}{$V(\Phi) \propto \Phi^{-2\beta}$, \\
$\xi(\Phi) \propto c_1 \Phi^{2\beta} + c_2 \Phi^{3\beta+1}$}}
& \multicolumn{3}{c|}{\parbox[c]{8cm}{
$V(\Phi)\propto c_1 \Phi^{-2\beta} + c_2 \Phi^{-\beta} + c_3 \Phi^{1-\beta}$, \\
$\xi(\Phi) \propto \left\{
\begin{array}{ll} \tilde c_1 \Phi^{\beta+1} + \tilde c_2 \Phi^{2\beta + 1}
+ \tilde c_3 \Phi^{2\beta + 2} \quad
& \left(\beta\neq -1,-1/2\right) \\
\tilde c_1 \ln \Phi + \tilde c_2 \Phi^{-1} + \tilde c_3 \Phi^{2\beta + 2} & \left(\beta=-1\right) \\
\tilde c_1 \Phi^{1/2} + \tilde c_2 \ln \Phi + \tilde c_3 \Phi^{2\beta + 2} & \left(\beta=-1/2\right)
\end{array}\right. $} } \\
\hline
Non-minimal Maxwell-Einstein gravity
& \multicolumn{1}{c|}{\parbox[c]{3.4cm}{$I(R) \propto R^{1+2h_0} \quad \left(\beta=1\right)$, \\
Not possible if $\beta>1$. }}
& \multicolumn{1}{c|}{Possible} & \multicolumn{1}{c|}{Not possible} & \multicolumn{1}{c|}{Possible} \\
\hline
\end{tabular}
}

\end{table}

\end{center}

Moreover, we have discussed several theoretical scenarios resolving
the finite-time future singularity. In particular, we have considered
the additional modification of the fluid with the inhomogeneous EoS or
that of the gravity action by the term which is relevant at the very
early (or very late) universe.
If the corresponding term is negligible at the current epoch,
such modification is always possible, at least, from the theoretical point
of view. Nevertheless, in order to check if the corresponding term is
realistic, its role at the very early/late universe should be confirmed
by using observational data.
Another scenario to remove the singularity is related with the account of
quantum effects (or even quantum gravity). However, the corresponding
consideration may give the preliminary results, at best, due to the absence
of the consistent theory of quantum gravity.

Furthermore, we have investigated the non-minimal Maxwell-Einstein
(or Maxwell-$F(R)$) gravity in the similar fashion.
We have studied the forms of the non-minimal gravitational coupling generating
the finite-time future singularities and the general conditions
for the non-minimal gravitational coupling in order that the finite-time
future singularities cannot emerge.
In addition, we have considered the influence of the non-minimal gravitational
coupling on the finite-time future singularities in modified gravity.
As a result, it has been shown that the non-minimal gravitational coupling
in Maxwell-modified gravity can remove the finite-time future singularities or
make the singularity stronger (or weaker).

The observational data of the current dark energy epoch still cannot
tell what is its exact nature: phantom, $\Lambda$CDM or
quintessence type.
It has been demonstrated in this paper that in some modified gravities with
the effective phantom or quintessence EoS the finite-time future singularity
can emerge. In this respect, the interpretation of the observational
data confirming (or excluding) the approach to the finite-time future
singularity is fundamentally important. From one side, it may clarify the
distant future of our universe. From another side, it may help to define the
evolution of the universe and the current value of the effective EoS
parameter: how close is $w$ to $-1$?

\section*{Acknowledgments}
We are grateful to M.~Sasaki for very helpful discussion of
related problems.
K.B. thanks the Kavli Institute for Theoretical Physics China at the Chinese
Academy of Sciences (KITPC-CAS)
for its very kind hospitality.
The work by S.D.O. was
supported in part by MEC (Spain) projects FIS2006-02842 and
PIE2007-50/023, RFBR grant 06-01-00609 and LRSS project N.2553.2008.2.
The work by S.N. is supported in part by the Ministry of Education,
Science, Sports and Culture of Japan under grant no.18549001 and Global
COE Program of Nagoya University provided by the Japan Society
for the Promotion of Science (G07).
The work by K.B. is supported in part by
National Tsing Hua University under Grant \#: 97N2309F1.

\appendix*
\section{\label{SecA1}}

We here consider Eq.~(\ref{eq:6.7}).
Let us now assume
\be
\label{II}
E_i = F_{0i}=\partial_0 A_i - \partial_i A_0 = 0\ ,
\ee
where $E_i$ is the electric field.
Then
\be
\label{III}
A_i = \partial_i \int dt A_0 + C_i\ .
\ee
Here $C_i$ does not depend on time $t$. Then we find
\be
\label{IV}
F_{ij} = \partial_i A_j - \partial_j A_i =  \partial_i C_j - \partial_j C_i\ .
\ee
Hence, the $F_{ij}$ and therefore the magnetic flux
$B_i \equiv \epsilon_{ijk} F_{jk}$
does not depend on time. Since the metric $g_{\mu\nu}$ and hence the
scalar
curvature $R$ do not depend on the spatial coordinates, Eq.~(\ref{eq:6.7})
reduces to the following form:
\be
\label{V}
\partial^{i} F_{ij} = \triangle C_j - \partial_j \partial^i C_i = 0\ .
\ee
The solution of (\ref{V}) is given by the constant $F_{ij}$ or constant
magnetic flux $B_i$.


\end{document}